\pdfoutput=1
\documentclass[11pt,a4paper]{article}
\usepackage{times}
\usepackage{latexsym}
\usepackage{amsmath}
\usepackage{dsfont}
\usepackage{graphicx}
\usepackage[ruled,vlined]{algorithm2e}
\usepackage{booktabs}
\usepackage{graphicx}
\PassOptionsToPackage{breaklinks}{hyperref}
\usepackage{xurl}
\usepackage[hyperref]{emnlp2020}
\SetKwComment{Comment}{$\triangleright$\ }{}

\usepackage{microtype}

\aclfinalcopy % Uncomment this line for the final submission
\title{Effective Distributed Representations for Academic Expert Search}

\author{Mark Berger \\
  University of Amsterdam \\
  Amsterdam, The Netherlands \\
  \texttt{mark@maberger.nl} \\\And
  Jakub Zavrel \\
  Zeta Alpha Vector \\
  Amsterdam, The Netherlands \\
  \texttt{zavrel@zeta-alpha.com} \\\And
  Paul Groth \\
  University of Amsterdam \\
  Amsterdam, The Netherlands \\
  \texttt{p.groth@uva.nl} \\}
\date{}

\begin{document}
\maketitle
\begin{abstract}
Expert search aims to find and rank experts based on a user's query. In academia, retrieving experts is an efficient way to navigate through a large amount of academic knowledge. Here, we study how different distributed representations of academic papers (i.e. embeddings) impact academic expert retrieval. We use the Microsoft Academic Graph dataset and experiment with different configurations of a document-centric voting model for retrieval. In particular, we explore the impact of the use of contextualized embeddings on search performance. We also present results for paper embeddings that incorporate citation information through retrofitting. Additionally, experiments are conducted using different techniques for assigning author weights based on author order. We observe that using contextual embeddings produced by a transformer model trained for sentence similarity tasks produces the most effective paper representations for document-centric expert retrieval. However, retrofitting the paper embeddings and using elaborate author contribution weighting strategies did not improve retrieval performance.
\end{abstract}

\section{Introduction}

To help navigate a large body of academic knowledge, it can be useful to identify expert individuals. Identifying such individuals may be useful to find collaborators \cite{10.5555/2035562.2035601, 10.1145/2147783.2147785, Sziklai2018HowTI}, to find paper reviewers \cite{silva2014research, price2017computational}, to find supervisors \cite{alarfaj-etal-2012-finding}, or to investigate literature in a certain domain. This process of identifying experts given a particular topic is called expert finding \cite{Balog2009}, expertise retrieval \cite{Goncalves2019}, or expert search. Expert search systems are information retrieval systems that can automatically rank candidate experts based on their expertise on a certain subject \citep{Husain2019}. In this study, we target the domain of retrieving academic experts based on papers they authored.

Given the central role of papers to defining expertise in this domain, we focus on document-centric expert search systems \cite{Balog2006}. These systems largely rely on statistical language modeling, topic modeling, or term frequency-based approaches to represent documents \cite{Goncalves2019, Husain2019}. Surprisingly, given the rapid advances in the field of contextualized text embeddings \cite{Wang2020}, little work has been done in applying these approaches to document representation for this task. We hypothesize that considering single words, which is common in the bag-of-words and probabilistic term-based approaches, may significantly reduce the system's ``understanding" of the underlying academic documents. To achieve a potentially deeper understanding of these papers, contextualized text embeddings could be used.

Thus, in this paper, we explore the impact of contextualized text embeddings on the performance of the expert search. Specifically, we make the following contributions:

\begin{itemize}
    \item a comparison of expert search performance using token-based (i.e. BERT \cite{Devlin2018}) and sentence-based (Sentence-BERT \cite{Reimers2019}) contextualized embeddings, non-contextualized embeddings (e.g. GloVe \cite{pennington2014glove}) and classic term frequency representations;
    \item measurement of the impact on performance when incorporating citation information into contextualized representations through \textit{retrofitting} \cite{Faruqui2015}; and \cite{Zhang2019retro}.
    \item a comparison of two different {\em strategies} for combining embeddings of the title and abstract of papers.
\end{itemize}
 
Additionally, all experiments are conducted using different techniques for assigning author weightings based on author order. Overall, this paper provides evidence for the efficacy of contextualized embeddings for the task of academic expert search. Note that this paper primarily focuses on investigating the performance of various contextualized embeddings and expert ranking aggregation methods within expert retrieval, and not on the entire retrieval process. Therefore, some aspects of neural information retrieval systems such as query understanding, query expansion, or re-ranking are out of the scope of this study.

Source code for the methods and data processing used in this paper can be found at GitHub\footnote{\url{https://github.com/mabergerx/SDP500_expert_search}}. The processed data used by our methods is available at \cite{markdataset2020}. 

The rest of this paper is organized as follows. We begin with a discussion of related work. Afterwards, the data used in this study is described. This is followed by a description of the various embeddings used and our approach to author ranking. Section \ref{evaluation-section} defines the evaluation and Section \ref{results} details its results. We, then, briefly describe a prototype implementation using these representations. Finally, we discuss the limitations of the work, potential future work and conclude. 

\section{Related work}
In this section, we introduce the primary paradigm for expert search. We then discuss work on voting models, document representations, and the use of text embedding techniques within expert search.

\paragraph{Probabilistic models}
A driving force behind expertise retrieval research was the launch of the TREC Enterprise Track in 2005 \cite{craswell2005overview}. This evaluation campaign led to the emergence of probabilistic models, in particular in the form of language models, as the primary paradigm for expertise retrieval. The core idea behind these approaches is to estimate a language model for each document and then rank the documents by the likelihood of the user query according to the language models \cite{Balog2009}.

\paragraph{Voting models}
We can see documents authored by experts as evidence for their expertise. A particular type of models, based on \textit{data fusion} methods that aggregate document scores into expert rankings, are called voting models \cite{Husain2019, Balog2012}.

Given a query, the retrieved documents are assumed to provide evidence about a possible ranking of the authors. 
This aggregation of the final author list can then be modelled as a voting process, where the document scores are aggregated into author scores \cite{macdonald2009voting, Macdonald2008, Macdonald2006, Macdonald2006_2}.

\paragraph{Paper embeddings}
Document-centric expert search systems rely on the documents to aggregate an expert ranking. However, effectively embedding longer documents is still an open research problem \cite{Beltagy2020, Zhang2016, Liu_docemb}.

Unsupervised document embedding techniques include Sent2Vec \cite{Pagliardini2017} and Doc2VecC \cite{chen2017efficient}, while supervised document embedding techniques include the Universal Sentence Encoder \cite{Cer2018} and InferSent \cite{Conneau2018}. Recently, the Longformer \cite{Beltagy2020} was proposed to embed even longer sequences of text than sentences. One research proposed evaluating various sentence encoding techniques in re-ranking of BM25-based research paper recommendations and found that the sentence encoding could be a beneficial method in addition to the BM25 retrieval, but not on its own \cite{Hassan2019}. Adding the BERT [CLS] token embedding into other ranking model's signal has been proposed and is shown to improve the underlying neural ranking architecture \cite{10.1145/3331184.3331317}.

As for the embedding of academic papers, most of the research focuses on learning the paper embeddings using linkage information and considers this a graph problem \cite{wang2016linked, Zhang2019, mai2018combining}.

\paragraph{Embedding expertise}
Given the amount of research on document embedding techniques, there has been surprisingly little attention given to the application of contextualized embedding techniques in the field of expertise retrieval. Three recent surveys and reviews on the field of expertise retrieval \cite{Goncalves2019, Husain2019, Lin2017} contained little to no information about the application of embedding techniques.

One of the first works to introduce this concept into expertise retrieval was Author2Vec \cite{Soumyajit2015}, which uses two models, the \emph{content-info model} and the \emph{link-info model} within the context of the co-authorship network. In the context-info model, the text of the written papers is represented using Paragraph2Vec \cite{Le2014}.

As briefly mentioned in the introduction, authors that cite each other can be considered having similar interests \cite{THO2007248, SHIBATA2008758}. Zhang \cite{Zhang2019retro} suggested using retrofitting in the domain of academic papers as a means of introducing this network information into the representation of a paper. Retrofitting is a concept introduced by Faruqui et al. \cite{Faruqui2015} which proposes the incorporation of the information from semantic lexicons such as WordNet into word embeddings.

\section{Data description}
The Microsoft Academic Graph (MAG) \cite{wang2020microsoft} was used at the primary data source. The data consists of over 200 million papers (titles \& abstracts) as well as a variety of metadata. We accessed the November 2018 snapshot of the MAG data through the Open Academic Graph initiative\footnote{\url{https://www.openacademic.ai/oag/}}, in particular the OAG v2 release. 

Due to the very large size of the MAG, we created a custom subset of the data that mainly consisted of Computer Science (CS) related papers. This domain allows us to interpret results better than other science domains.

Our approach in extracting Computer Science (CS) related papers was to take the 113.864 paper titles obtained from arXiv\footnote{\url{https://arxiv.org}} - a widely used preprint server - and do an exhaustive title matching on the full MAG dataset. This search resulted in 29.237 exact title matches, which corresponds to 26,6\% of the arXiv data. This set provided us with a substantial initial seed of papers to extract more CS papers from the MAG data.

To allow retrofitting later in the process and create a larger dataset, we expanded this set with the references of all the 29.237 papers, which resulted in a set of 221.347 papers. These references were retrieved by accessing the \textit{references} field of each of the 29.237 paper in the MAG data. Note that these references are not necessarily always complete: some cited articles may not be present in our data due to incompleteness of the source MAG data.

From these 221.347 papers, we then performed bounded stratified sampling for the authors to retrieve a subset of 5.000 authors who are representative of both highly-, medium- and less prolific author populations. The full sampling method is described in Algorithm \ref{sampling-strategy} in the Algorithms appendix.

This set of 5.000 authors served as a starting point for a second, final round of data retrieval. For these authors, we retrieved all their papers and references, resulting in a set of 127.716 papers, which included authors of the referenced papers. For all these new authors, we collected the metadata from the MAG authors dataset and aggregated this information into a single final authors dataset. 
The reason for expanding the set of authors beyond the 5.000 sampled authors is that a larger pool of papers is beneficial for the retrieval due to the larger search space. 

For all titles and abstracts in our dataset, we performed data cleaning. Specifically, (corpus-specific) stopwords were removed, redundant whitespace and Unicode characters were both normalized. URLs and e-mails were removed.

\section{Paper embedding methodology}
In this section, we describe the paper embedding techniques employed and discuss our approach to embedding indexing and search, as well as retrofitting embeddings.

\subsection{Embedding techniques}
\label{embedding-techniques-method}
Various approaches have been used to embed the papers. We divide our approaches into the custom contextual approach and the baseline approaches. In all our baseline approaches, we use the concatenation of the title and the abstract to represent the paper.

\paragraph{Custom contextual approach}
\label{method-contextual-approach}

We use the title and the abstract as representative texts for a paper. Although the title and abstract of a paper are both relevant representations, they may contain information that differs in importance and granularity. In order to capture the possible semantic weight differences between the title and the abstract, we deploy two different embedding combination strategies: the \textbf{merge strategy} and the \textbf{separate strategy}.

In the merge strategy, we assume that the semantic weights of the title sentence and the abstract sentences are equal. That means that we want to take the average over the title- and abstract sentence embeddings without assigning any extra weight to neither. 
A detailed specification is given in the Algorithm \ref{merged-strategy} in the appendix.

In the separate strategy, we do want to differentiate between the title and the abstract. In particular, we want to assign more weight to the title than to the individual abstract sentences. We achieve this by first computing the average abstract embedding and then taking the average between that and the title embedding. 
A detailed specification is given in the Algorithm \ref{separate-strategy} in the appendix.

We refer to the actual text embedding model as the \textit{embedder}. The \textit{embedder} of choice is Sentence-BERT \cite{Reimers2019}, which is specifically designed for producing meaningful sentence-level embeddings, suited for Semantic Textual Similarity (STS). Specifically, we make use of the RoBERTa-base model fine-tuned on the combination of NLI datasets, and then further fine-tuned on the STS benchmark training set \footnote{\url{https://github.com/UKPLab/sentence-transformers}}. 

\paragraph{Baseline approach: Latent Semantic Indexing (LSI) \cite{deerwester1990indexing}}
\label{method-baseline-approach}

TF-IDF vectors with applied singular value decomposition. For our experiments, we chose to set the LSI vector's dimensionality to 768 dimensions, the same as the Sentence-BERT embedding dimensionality.

\paragraph{Baseline approach: BERT- and GloVe pooling}
\label{method-baseline-approach-2}

To provide a comparison between specifically tuned for document-level representations Sentence-BERT and conventional pooling document embedding techniques, we produced paper embeddings by averaging both BERT and GloVe token embeddings. In both averaging operations, we perform double pooling: first, all tokens within each sentence are embedded and averaged into a single sentence embedding, and then these sentence embeddings are once again averaged into a single paper embedding.
The details of this embedding process are shown in Algorithm \ref{baseline-bert-glove-strategy} in the appendix and is identical for both BERT and GloVe embeddings. 
For both BERT (\textit{bert-base-uncased}) and GloVe embedding calculations, we used the \textit{Flair} \cite{akbik2018coling} Python library. 

\subsection{Retrofitting}
\label{retrofitting-section}

Authors that cite each other can be considered as having similar interests. In the context of having a semantic representation of expertise, it could be helpful to “expand" a paper embedding to broaden the expertise scope of the author beyond a particular paper. To achieve this broadening, we use a technique called retrofitting, which introduces network information into the embeddings.

Inspired by \cite{Zhang2019retro}, we adapt the original implementation\footnote{\label{retrogit}\url{https://github.com/mfaruqui/retrofitting}} of retrofitting \cite{Faruqui2015} to work with academic papers that have been contextually embedded. 
% The process of data formatting for retrofitting is detailed in Algorithm \ref{retro-lexicon-creation} in the appendix. 
The retrofitting process is performed for ten iterations. Algorithm \ref{retro-algo-figure} in the appendix shows the details about the algorithm.

\subsection{Embedding storage and search}
\label{section-embedding-storage-search}

We chose the FAISS \cite{JDH17} library by Facebook for our indexing purposes. It is optimized for memory usage and speed and can handle a large number of vectors.

For our embeddings, we chose the \textit{IndexHNSWFlat} index \footnote{\url{https://github.com/facebookresearch/faiss/wiki/Faiss-indexes}}. We use cosine similarity as the measure of similarity between the query embedding $\vec{Q}$ and any of the indexed embeddings $\vec{V}$.

\section{Author ranking via voting}
\label{voting-data-fusion-method}
From the FAISS index we can, given a query, retrieve top \textit{N} similar papers. To produce a final author ranking, we adopt a {\em voting model} based approach. 

We can consider the retrieved paper results as the “expertise evidence” for the authors of these papers. A range of different voting approaches based on data fusion techniques has been proposed \cite{Macdonald2008, afzal2011expertise, 10.5555/2385736.2385738} to produce an author ranking given the documents.

Each retrieved document $d$ from the set of retrieved documents $R(Q)$ has an associated similarity score $s(d, Q)$ to it, with regard to the query $Q$. We can then combine these document scores into aggregated author scores using the \textit{ExpCombSUM (eCS)} data fusion function \cite{macdonald2009voting}:

\begin{equation}
    eCS(C, Q) = \sum\limits_{d \in R(Q) \cap D_C} e^{(s(d,Q)))}
\end{equation}

where $C$ is a candidate expert, $D_C$ is the set of documents associated with candidate C.

This algorithm \cite{Macdonald2008, macdonald2009voting}, assumes that each document produces a static score per related author. In the case of academic papers, that is not the case, as most papers have multiple authors. These authors mostly have a different level of involvement in a particular paper and, therefore, may have a different vote produced by the document, depending on their authorship role. Recent research has shown that because the research is increasingly more interdisciplinary, evaluating authors based on their rank within the order of authors is becoming increasingly difficult \cite{Jnior2017PatternsOA}. Therefore, it could be valuable to assign different weights to different authors of the same document. To the best of our knowledge, no previous work has been done on exactly defining weights on the authorship scores within a voting model. We define four different weighting strategies:

\begin{enumerate}
    \item \textbf{Binary weighting}. Each author gets the full score for a document. This strategy assumes that each author contributed equally.
    \item \textbf{Uniform weighting}. Each author gets $\frac{\textit{fullScore}}{\textit{\# authors}}$ for a document. This strategy assumes that each author contributed equally but does normalize the score by the number of authors.
    \item \textbf{Descending weighting}. The first author gets the full document score. Each following author gets \ $\textit{fullScore} * decayFactor$, where $decayFactor$ starts at 0.8 and decreases with 0.2 for each consecutive author. This strategy assumes that the authors are listed in descending involvement order.
    \item \textbf{Parabolic weighting}. The first and last author get the full document score. All authors in between follow the \textbf{descending weighting}. This strategy assumes that the first author is similar to the descending weighting, but also takes the possible importance of the last author as the project supervisor.
\end{enumerate}

Using these data fusion approaches, a fairness problem may occur: highly prolific authors, that may be associated with many documents (for instance, because they are the head of a lab) may receive an unfairly large number of votes, which does not necessarily indicate their expertise. Candidate length normalization is proposed to deal with this unfairness, just as document length normalization is often performed in document retrieval systems \cite{macdonald2009voting}.

The use of a classical document normalization technique based on the \textit{Divergence From Randomness framework} \cite{amati2003probabilistic} is proposed \cite{macdonald2009voting}, and has the following formula:

\begin{equation}
    s_{N}(C, Q) = s(C, Q) \cdot log_2(1 + \alpha \cdot \frac{aL}{lP})
\end{equation}

where $\alpha$ is a hyperparameter controlling the amount of normalization, $aL$ is the average amount of publications, and $lP$ is the length of the profile of the candidate $C$. The lower the $\alpha$ parameter is, the more less prolific authors are boosted, and the more highly prolific authors are suppressed.

In practice, we discovered that because of the nature of our dataset, if we apply the above normalization technique, many authors from the long tail are retrieved, even when we use high $\alpha$ values. Therefore, we experimented with introducing another term to the equation: $\beta$, which serves as a profile length “booster":

\begin{equation}
    s_{N}(C, Q) = s(C, Q) \cdot log_2(1 + \alpha \cdot \frac{aL}{lP + \beta})
\end{equation}

While it does introduce bias and eases the normalization, in our case, it provided an extra parameter to tune and resulted in a better mix of well- and lesser-known authors.

\section{Evaluation methods}
\label{evaluation-section}
To evaluate the different retrieval strategies, we developed a method which uses the \textit{field of work} tags present in the MAG dataset for the authors as a proxy for evaluating the relevance of an author. Because we use a document-centric retrieval strategy which uses strictly only the embeddings of the paper's title and abstract, the \textit{field of work} tags are not used in the retrieval process, allowing us to use these tags in the evaluation process. 

% \subsection{Query test set}
The query test set for our research was selected from the full distribution of author tags. The final query test set contained a hundred Computer Science related queries. The set is available in Table \ref{tab:test_queries} in the appendix. 

\subsection{Relevance metrics}
\label{eval-binary-relevance}
Before we can use any of the existing binary information retrieval metrics, we need to define a notion of \textit{relevance} of an author given a query. 

Based on the \textit{field of work} tags, we define two relevance checks:

\begin{enumerate}
    \item \textbf{Exact topic query evaluation \cite{Brochier2018}}. This approach takes a description of a topic and uses it directly as a query. The experts associated with that topic are then the ground truth list of candidates to be retrieved. In our case, the \textit{field of work} tags of the authors are used as queries, and the retrieved authors are labelled relevant if they have that tag.
    \item \textbf{Approximate topic query evaluation}. Sometimes, a query retrieves authors who have an incomplete \textit{field of work} tag list or have tags which are very similar to the query but do not exactly match it. For instance, author $A$ may have \textit{“automatic summarization"} in their tags list, while the query was \textit{“automatic text summarization"}. This author is clearly highly relevant to the query but would be labelled as irrelevant by the exact topic query evaluation method. Therefore, we introduce a fuzzy relevance checking method which, given a query, calculates the cosine similarity between the query embedding and each of the author's tags. If any of the similarities are higher than a chosen threshold, then we deem the author relevant. 
\end{enumerate}

Once we can label each retrieved author as relevant or not relevant, we can use different evaluation metrics. We evaluate our system by using three binary relevance metrics, all measured @N and using both the exact and approximate topic query evaluation: \textbf{Mean Reciprocal Rank} ($MRR@N_{exact}$, $MRR@N_{approx}$), \textbf{Mean Precision @ N} ($MP@N_{exact}$, $MP@N_{approx}$), and  \textbf{Mean Average Precision} ($MAP@N_{exact}$, $MAP@N_{approx}$)

% \subsection{Gain based relevance}
% \label{eval-gain-relevance}

Some authors are more relevant than others. To incorporate this into our evaluation, we also use the \textbf{Normalized discounted cumulative gain} (nDCG@N) score, which is sensitive to the position of the relevant items in the produced ranking.

To produce the nDCG scores, we first need to have the score for the ideal ranking given a query, IDCG. To calculate the IDCG scores, for each query in our test set, we created a mapping between the query and the corresponding top authors. Each author in such mapping got a relevance label in relation to the query. In our implementation of the author's relevance, we use the citations of the relevant papers of an author as a proxy for the expertise. Although any expertise measure of an author is not fully objective, and many factors seem to have effect on the citation activity \cite{10.1145/2063576.2063757}, we chose the citation counts of the papers as a proxy for expertise for the following reasons:
\begin{itemize}
    \item This measure is explainable.
    \item It prevents highly prolific but rarely cited authors to be labeled as highly relevant for multiple topics just based on their output
\end{itemize}

Given this final query-to-expert mapping, we precalculated the IDCG@10 score for each of the test queries in our dataset, so we could later calculate the nDCG@10 score per query. Once we calculated all the scores for our test set, we can take the average of those scores to have a single nDCG@10 score for our current system.

\section{Results}
\label{results}

In this section, first the overall quantitative evaluation results are discussed and then we zoom in on the performance of retrofitting and author contribution weighting. 

\subsection{Voting model results}
% \usepackage{graphicx}
\begin{table*}[!h]
\centering
\resizebox{\textwidth}{!}{%
\begin{tabular}{llllllll}
\hline
\textbf{PAPER EMBEDDING KIND} &
  \textbf{DATA FUSION TECHNIQUE} &
  \textbf{\begin{tabular}[c]{@{}l@{}}MRR@10 \\ EXACT\end{tabular}} &
%   \textbf{\begin{tabular}[c]{@{}l@{}}MRR@10 \\ APPRX\end{tabular}} &
  \textbf{\begin{tabular}[c]{@{}l@{}}MAP@10 \\ EXACT\end{tabular}} &
%   \textbf{\begin{tabular}[c]{@{}l@{}}MAP@10 \\ APPRX\end{tabular}} &
  \textbf{\begin{tabular}[c]{@{}l@{}}MP@10\\ EXACT\end{tabular}} &
%   \textbf{\begin{tabular}[c]{@{}l@{}}MP@10\\ APPRX\end{tabular}} &
  \textbf{\begin{tabular}[c]{@{}l@{}}MP@5 \\ EXACT\end{tabular}} &
%   \textbf{\begin{tabular}[c]{@{}l@{}}MP@5 \\ APPRX\end{tabular}} &
 \textbf{\begin{tabular}[c]{@{}l@{}}NDCG \\ @10\end{tabular}} &
 \textbf{\begin{tabular}[c]{@{}l@{}}NDCG \\ @5\end{tabular}} \\ \hline
%   \textbf{NDCG@10} &
%   \textbf{NDCG@5} \\ \hline
 LSI &
  $expCombSUM_{uniform}$ &
  0.75 &
%   0.864 &
  0.399 &
%   0.603 &
  0.462 &
%   0.662 &
  0.496 &
%   0.696 &
  0.39 &
  0.42 \\ \hline
 &
  $expCombSUM_{binary}$ &
  0.753 &
%   0.844 &
  0.428 &
%   0.626 &
  0.491 &
%   0.693 &
  0.512 &
%   0.708 &
  0.41 &
  0.44 \\ \hline
 &
  $expCombSUM_{descending}$ &
  0.755 &
%   0.857 &
  0.411 &
%   0.603 &
  0.476 &
%   0.668 &
  0.484 &
%   0.684 &
  0.39 &
  0.42 \\ \hline
 &
  $expCombSUM_{parabolic}$ &
  0.763 &
%   0.864 &
  0.392 &
%   0.587 &
  0.457 &
%   0.651 &
  0.482 &
%   0.68 &
  0.39 &
  0.42 \\ \hline
Average pooled BERT &
  $expCombSUM_{uniform}$ &
  0.554 &
%   - &
  0.117 &
%   - &
  0.198 &
%   - &
  0.206 &
%   - &
  0.1 &
  0.11 \\ \hline
 &
  $expCombSUM_{binary}$ &
  0.558 &
%   - &
  0.12 &
%   - &
  0.203 &
%   - &
  0.222 &
%   - &
  0.12 &
  0.13 \\ \hline
 &
  $expCombSUM_{descending}$ &
  0.556 &
%   - &
  0.117 &
%   - &
  0.2 &
%   - &
  0.208 &
%   - &
  0.11 &
  0.12 \\ \hline
 &
  $expCombSUM_{parabolic}$ &
  0.56 &
%   - &
  0.107 &
%   - &
  0.18 &
%   - &
  0.218 &
%   - &
  0.1 &
  0.12 \\ \hline
\begin{tabular}[c]{@{}l@{}}Average pooled GloVe\end{tabular} &
  $expCombSUM_{uniform}$ &
  0.626 &
%   - &
  0.272 &
%   - &
  0.369 &
%   - &
  0.398 &
%   - &
  0.27 &
  0.29 \\ \hline
 &
  $expCombSUM_{binary}$ &
  0.672 &
%   - &
  0.304 &
%   - &
  0.402 &
%   - &
  0.414 &
%   - &
  0.3 &
  0.31 \\ \hline
 &
  $expCombSUM_{descending}$ &
  0.661 &
%   - &
  0.286 &
%   - &
  0.383 &
%   - &
  0.398 &
%   - &
  0.28 &
  0.3 \\ \hline
 &
  $expCombSUM_{parabolic}$ &
  0.676 &
%   - &
  0.254 &
%   - &
  0.33 &
%   - &
  0.392 &
%   - &
  0.25 &
  0.28 \\ \hline
Merged Sentence-BERT &
  $expCombSUM_{uniform}$ &
  0.834 &
%   \textbf{0.903} &
  0.419 &
%   0.586 &
  0.491 &
%   0.658 &
  0.528 &
%   0.684 &
  0.43 &
  0.47 \\ \hline
 &
  $expCombSUM_{binary}$ &
  0.83 &
%   0.893 &
  0.437 &
%   0.606 &
  0.509 &
%   0.677 &
  0.546 &
%   0.708 &
  0.42 &
  0.46 \\ \hline
 &
  $expCombSUM_{descending}$ &
  0.818 &
%   \textbf{0.903} &
  0.419 &
%   0.59 &
  0.493 &
%   0.657 &
  0.526 &
%   0.688 &
  0.41 &
  0.46 \\ \hline
 &
  $expCombSUM_{parabolic}$ &
  0.812 &
%   0.869 &
  0.4 &
%   0.555 &
  0.484 &
%   0.634 &
  0.508 &
%   0.666 &
  0.41 &
  0.45 \\ \hline
\textbf{Separate Sentence-BERT} &
  $expCombSUM_{uniform}$ &
  0.838 &
%   0.875 &
  0.495 &
%   0.673 &
  0.572 &
%   0.744 &
  0.616 &
%   0.764 &
  0.53 &
  0.58 \\ \hline
 &
  $\mathbf{expCombSUM_{binary}}$ &
  0.837 &
%   0.892 &
  0.518 &
%   \textbf{0.69} &
  \textbf{0.59} &
%   \textbf{0.751} &
  \textbf{0.626} &
%   \textbf{0.784} &
  \textbf{0.54} &
  \textbf{0.6} \\ \hline
 &
  \begin{tabular}[c]{@{}l@{}}$Norm(expCombSUM_{binary})$\\ $\beta =0$ and $\alpha=1$\end{tabular} &
  0.619 &
%   0.678 &
  0.174 &
%   0.284 &
  0.282 &
%   0.42 &
  0.272 &
%   0.398 &
  0.15 &
  0.14 \\ \hline
 &
  \begin{tabular}[c]{@{}l@{}}$Norm(expCombSUM_{binary})$\\ $\beta =0$ and $\alpha=1000$\end{tabular} &
  0.694 &
%   0.737 &
  0.218 &
%   0.331 &
  0.318 &
%   0.452 &
  0.324 &
%   0.456 &
  0.16 &
  0.17 \\ \hline
 &
  \begin{tabular}[c]{@{}l@{}}$Norm(expCombSUM_{binary})$\\ $\beta =10$ and $\alpha=1000$\end{tabular} &
  0.777 &
%   0.807 &
  0.293 &
%   0.409 &
  0.381 &
%   0.509 &
  0.406 &
%   0.542 &
  0.22 &
  0.25 \\ \hline
 &
  \begin{tabular}[c]{@{}l@{}}$Norm(expCombSUM_{binary})$\\ $\beta = 50$ and $\alpha=1000$\end{tabular} &
  0.769 &
%   0.81 &
  0.362 &
%   0.49 &
  0.455 &
%   0.589 &
  0.466 &
%   0.602 &
  0.31 &
  0.33 \\ \hline
 &
  \begin{tabular}[c]{@{}l@{}}$Norm(expCombSUM_{binary})$\\ $\beta =1000$ and $\alpha=1000$\end{tabular} &
  0.813 &
%   0.853 &
  0.404 &
%   0.548 &
  0.491 &
%   0.638 &
  0.52 &
%   0.658 &
  0.37 &
  0.4 \\ \hline
 &
  $expCombSUM_{descending}$ &
  0.839 &
%   0.894 &
  0.501 &
%   0.666 &
  0.581 &
%   0.737 &
  0.612 &
%   0.758 &
  0.52 &
  0.58 \\ \hline
 &
  $expCombSUM_{parabolic}$ &
  0.819 &
%   0.878 &
  0.486 &
%   0.654 &
  0.565 &
%   0.729 &
  0.592 &
%   0.742 &
  0.52 &
  0.56 \\ \hline
Retrofitted merged Sentence-BERT &
  $expCombSUM_{uniform}$ &
  0.792 &
%   0.839 &
  0.384 &
%   0.551 &
  0.45 &
%   0.61 &
  0.482 &
%   0.638 &
  0.38 &
  0.42 \\ \hline
 &
  $expCombSUM_{binary}$ &
  0.83 &
%   0.865 &
  0.404 &
%   0.563 &
  0.467 &
%   0.618 &
  0.496 &
%   0.652 &
  0.39 &
  0.44 \\ \hline
 &
  $expCombSUM_{descending}$ &
  0.813 &
%   0.843 &
  0.39 &
%   0.551 &
  0.454 &
%   0.609 &
  0.486 &
%   0.644 &
  0.38 &
  0.42 \\ \hline
 &
  $expCombSUM_{parabolic}$ &
  0.775 &
%   0.847 &
  0.38 &
%   0.542 &
  0.445 &
%   0.598 &
  0.474 &
%   0.638 &
  0.38 &
  0.4 \\ \hline
Retrofitted separate Sentence-BERT &
  $expCombSUM_{uniform}$ &
  0.821 &
%   0.893 &
  0.51 &
%   0.658 &
  0.577 &
%   0.716 &
  0.606 &
%   0.732 &
  0.5 &
  0.54 \\ \hline
 &
  $expCombSUM_{binary}$ &
  \textbf{0.841} &
%   0.893 &
  \textbf{0.519} &
%   0.661 &
  0.584 &
%   0.718 &
  0.616 &
%   0.75 &
  0.51 &
  0.54 \\ \hline
 &
  $expCombSUM_{descending}$ &
  0.831 &
%   0.895 &
  0.505 &
%   0.647 &
  0.569 &
%   0.702 &
  0.61 &
%   0.734 &
  0.49 &
  0.54 \\ \hline
 &
  $expCombSUM_{parabolic}$ &
  0.808 &
%   0.863 &
  0.509 &
%   0.658 &
  0.583 &
%   0.724 &
  0.596 &
%   0.732 &
  0.5 &
  0.53 \\ \hline
\end{tabular}%
}
\caption{\small{Results for the voting model author retrieval strategy. The best results are formatted in bold.}}
\label{tab:results_voting_author_retrieval_paper}
\end{table*}
The results of the voting model approach are presented in Table \ref{tab:results_voting_author_retrieval_paper}. Here we only present the exact topic query evaluation results, as we observe that approximate topic query evaluation results correspond to the exact results but are overall higher. In particular, for MRR, the approximate results are 0.06 higher on average; for MAP, the approximate results are 0.16 higher; for MP@10 the approximate results are 0.15 higher; and finally the MP@5 approximate results are, again, 0.15 higher. The full results table with the approximate query evaluation results included, are presented in Table \ref{tab:results_voting_author_retrieval_full} in the Appendix.

We can observe that the LSI baseline produces strong results, outperforming both embedding pooling baselines. From the four used author contribution weighting schemes, the binary score weighting is the best performing weighting. However, the overall performance difference between the weightings is quite small.

The best performing configuration is the separate embedding strategy with the binary distributed paper scores. We see that normalizing the \textit{expCombSUM} function with a low $\alpha$ and $\beta=0$ leads to steep decrease in performance. With higher $\alpha$ and $\beta$, the performance goes up but can be explained by “cancelling out" the normalization effect. One of the reasons for this bad performance could be that the dataset contains many authors from the long tail; some of the authors naturally may have worse metadata resulting in missing expertise tags. Moreover, for lesser-known authors in the MAG, we have encountered a problem that the profiles get deleted or get different author ids, which also corrupts the metadata and the results.

Retrofitting the embeddings did not improve the results, except for two evaluation metrics: $MRR@10_{exact}$ and $MAP@10_{exact}$ in case of the retrofitted separate embeddings. 

\subsection{Performance of retrofitting}
\label{results-performance-of-retrofitting}
Retrofitting the embeddings did not improve our retrieval performance with the exception of two metrics. One of the explanations could be that the relatively small size of our dataset can hurt the retrofitting process. Combined with the variance in the number of neighbours per paper, some of the resulting retrofitted embeddings might be driven away too much from the original embedding, while other embeddings are not modified “enough".

\subsection{Effect of different author contribution weightings}
\label{results-author-contrib-explanation}
Throughout the experiment, we observed that the binary author contribution weighting performs the best. Therefore, we can conclude that, for our voting model configuration, introducing elaborate author contribution combinations does not improve the retrieval performance.

\section{Prototype implementation}
We implemented our model int a prototype, which consists of a REST API, where the users can, given a search query, look for \textit{N} most relevant experts. Each individual expert representation is contained within a JSON object which contains not only the author's name and MAG id, but also the authors affiliation information (retrieved using GRID \footnote{\url{https://www.grid.ac/}}), the list of papers which voted for the author and their corresponding document scores in relation to the query, and additional author information from WikiData \cite{vrandevcic2014wikidata}.

\section{Discussion \& Future work}
This study faced multiple challenges regarding data quality, data freshness, embedding strategy considerations, and the retrieval base on embeddings. In this section, we discuss these issues.

{\em Data completeness and variability.}
\label{data-completeness-discussion}
The \textit{field-of-work} tags used in the evaluation are not always complete and are subject to the specific format used by the MAG. In addition, for some authors, no tags are present in our snapshot of the MAG data. This has an effect on the evaluation performance by introducing both false positives and -negatives into the relevancy determining process. Overall, our system would profit significantly from a larger pool of papers and authors.

{\em Shifting expertise.} \label{shifting-exp-discussion}
Many authors have varying interests during their academic career. Our system is inherently a snapshot of their academic activity: we only search and aggregate within a bounded set of papers. Introducing temporal aspect into the author search, which would intelligently account for shifting expertise could improve the retrieval results.

{\em Representing expertise domains.}
Many authors are not experts in just one niche field, but rather are knowledgeable about a pretty broad field of science, with more in-depth knowledge about a few specific sub-fields. Clustering within the author expertise to find expertise sub-clusters might help to create more nuanced author representations by taking the cluster centroids as the individual author embeddings. This approach, while interesting, requires more data, as clustering within the papers of one author requires having a significant amount of papers per author.

{\em Performance of individual retrieval strategies.} Different types of embeddings may perform well in different scenarios. For example, retrofitting of the paper embeddings leads to the “widening" of the semantic representation of a paper into the direction of its neighbours. For example, retrofitted embeddings may perform better for more broad/general queries and worse on more specific queries. The same applies for normalization in the voting model: retrieving less prolific authors may be beneficial for some user's needs, while it technically hurts the quantitative performance of the system.

{\em Document pooling strategies.} In the two paper embedding strategies, we perform pooling over multiple sentence embeddings. While these strategies seem to work well to represent the papers for our task, it would still be interesting to use new, state-of-the-art approaches for embedding longer texts, such as the Longformer \citep{Beltagy2020}, to avoid using any pooling strategies and loosing semantic value. Techniques like the recently introduced SPECTER \cite{specter_cohan_2020} could also be used to produce better citation-informed document embeddings. Finally, we could employ the SciBERT \cite{Beltagy2019SciBERT} model in our baseline pooling strategies or even fine-tune a SciBERT model on longer sequences, similarly to Sentence-BERT, as it is better suited for academic texts.

{\em Graph embeddings.} In our approach, we used retrofitting to introduce citation network information into our initial paper embeddings. However, we could also go a step further and use graph embedding techniques to create native graph-based embeddings for our papers \cite{mai2018combining, wang2016linked, Zhang2019}.

\section{Conclusion}
In this study, we investigated different approaches for embedding academic papers and using the embeddings in an expert search task. 

Overall, we found that Transformer-based contextual text embeddings work well on the domain of academic papers. By using the Sentence-BERT model trained on NLI and SNS tasks, we outperformed the strong LSI baseline often employed in information retrieval systems on all ten evaluation metrics. We also outperformed the baseline strategies of average pooled BERT and GloVe embeddings.

Employing a weighted embedding combination strategy to represent a paper can however be valuable, as we found that using a separate embedding combination strategy outperformed the ``default" merged strategy on nine out of ten metrics.

We hypothesized that enriching the paper embeddings with citation information, in a process called retrofitting, could improve improve retrieval performance. Our experiments did not confirm this hypothesis, as non-retrofitted embeddings performed better in our task on all but two evaluation metrics.

Finally, we employed various data fusion techniques to convert the top \textit{N} retrieved papers given a query into a ranking of authors. A voting model was used, where each document served as evidence for the corresponding author's expertise. We investigated whether using author contribution weighting strategies within the voting process would improve expertise retrieval. We observed no performance gain over the ``default" binary strategy.

Given the direction of this study, we think that the most suitable application areas for the methodology we proposed are reviewer finding, supervisor finding and investigating literature on a topic. The reasoning behind this is that finding a collaborator may require and involve more sophisticated information about the institution, availability and current field of interest of the found experts. The proposed three areas, however, allow for less specificity and can better benefit from the improved retrieval.

While research in the field of expertise retrieval is not as active in the second half of the 2010s as it was in the first half, the area of text representations and retrieval has seen dramatic improvements. This study was an effort to apply these new techniques into the field of expertise retrieval and has shown that substantial improvements can be made over the existing retrieval algorithms. We hope that this study can contribute to a new research wave within the field of (academic) expertise retrieval.

\section*{Acknowledgements}
We would like to thank the reviewers for their insightful commentary. Additionally, we would like to thank dr. Wouter Weerkamp from Zeta Alpha for his valuable feedback and assessment of the work.

\nocite{*} %

\bibliography{main_paper}

\begin{thebibliography}{93}
\expandafter\ifx\csname natexlab\endcsname\relax\def\natexlab#1{#1}\fi

\bibitem[{Afzal and Maurer(2011)}]{afzal2011expertise}
Muhammad~Tanvir Afzal and Hermann~A Maurer. 2011.
\newblock \href {https://doi.org/10.3217/jucs-017-11-1529} {Expertise
  recommender system for scientific community.}
\newblock \emph{J. UCS}, 17(11):1529--1549.

\bibitem[{Akbik et~al.(2018)Akbik, Blythe, and Vollgraf}]{akbik2018coling}
Alan Akbik, Duncan Blythe, and Roland Vollgraf. 2018.
\newblock Contextual string embeddings for sequence labeling.
\newblock In \emph{{COLING} 2018, 27th International Conference on
  Computational Linguistics}, pages 1638--1649.

\bibitem[{Alarfaj et~al.(2012{\natexlab{a}})Alarfaj, Kruschwitz, Hunter, and
  Fox}]{alarfaj-etal-2012-finding}
Fawaz Alarfaj, Udo Kruschwitz, David Hunter, and Chris Fox. 2012{\natexlab{a}}.
\newblock \href {https://www.aclweb.org/anthology/N12-2001} {Finding the right
  supervisor: Expert-finding in a university domain}.
\newblock In \emph{Proceedings of the {NAACL} {HLT} 2012 Student Research
  Workshop}, pages 1--6, Montr{\'e}al, Canada. Association for Computational
  Linguistics.

\bibitem[{Alarfaj et~al.(2012{\natexlab{b}})Alarfaj, Kruschwitz, Hunter, and
  Fox}]{10.5555/2385736.2385738}
Fawaz Alarfaj, Udo Kruschwitz, David Hunter, and Chris Fox. 2012{\natexlab{b}}.
\newblock \href {https://www.aclweb.org/anthology/N12-2001} {Finding the right
  supervisor: Expert-finding in a university domain}.
\newblock In \emph{Proceedings of the 2012 Conference of the North American
  Chapter of the Association for Computational Linguistics: Human Language
  Technologies: Student Research Workshop}, NAACL HLT ’12, page 1–6, USA.
  Association for Computational Linguistics.

\bibitem[{Amati(2003)}]{amati2003probabilistic}
G~Amati. 2003.
\newblock \emph{Probabilistic Models for Information Retrieval based on
  Divergence from Randomness. University of Glasgow, UK}.
\newblock Ph.D. thesis, PhD Thesis.

\bibitem[{Ammar et~al.(2018)Ammar, Groeneveld, Bhagavatula, Beltagy, Crawford,
  Downey, Dunkelberger, Elgohary, Feldman, Ha et~al.}]{ammar2018construction}
Waleed Ammar, Dirk Groeneveld, Chandra Bhagavatula, Iz~Beltagy, Miles Crawford,
  Doug Downey, Jason Dunkelberger, Ahmed Elgohary, Sergey Feldman, Vu~Ha,
  et~al. 2018.
\newblock Construction of the literature graph in semantic scholar.
\newblock \emph{arXiv preprint arXiv:1805.02262}.

\bibitem[{Arora et~al.(2019)Arora, Liang, and
  Ma}]{2b29624234f9441cab6dd9c918e86ab7}
Sanjeev Arora, Yingyu Liang, and Tengyu Ma. 2019.
\newblock A simple but tough-to-beat baseline for sentence embeddings.
\newblock 5th International Conference on Learning Representations, ICLR 2017 ;
  Conference date: 24-04-2017 Through 26-04-2017.

\bibitem[{Balog et~al.(2006)Balog, Azzopardi, and {De Rijke}}]{Balog2006}
Krisztian Balog, Leif Azzopardi, and Maarten {De Rijke}. 2006.
\newblock \href {https://doi.org/10.1145/1148170.1148181} {{Formal models for
  expert finding in enterprise corpora}}.
\newblock \emph{Proceedings of the Twenty-Ninth Annual International ACM SIGIR
  Conference on Research and Development in Information Retrieval},
  2006:43--50.

\bibitem[{Balog et~al.(2009)Balog, Azzopardi, and de~Rijke}]{Balog2009}
Krisztian Balog, Leif Azzopardi, and Maarten de~Rijke. 2009.
\newblock \href {https://doi.org/10.1016/j.ipm.2008.06.003} {{A language
  modeling framework for expert finding}}.
\newblock \emph{Information Processing and Management}, 45(1):1--19.

\bibitem[{Balog et~al.(2007)Balog, Bogers, Azzopardi, de~Rijke, and van~den
  Bosch}]{10.1145/1277741.1277836}
Krisztian Balog, Toine Bogers, Leif Azzopardi, Maarten de~Rijke, and Antal
  van~den Bosch. 2007.
\newblock \href {https://doi.org/10.1145/1277741.1277836} {Broad expertise
  retrieval in sparse data environments}.
\newblock In \emph{Proceedings of the 30th Annual International ACM SIGIR
  Conference on Research and Development in Information Retrieval}, SIGIR
  ’07, page 551–558, New York, NY, USA. Association for Computing
  Machinery.

\bibitem[{Balog et~al.(2012)Balog, Fang, {De Rijke}, Serdyukov, and
  Si}]{Balog2012}
Krisztian Balog, Yi~Fang, Maarten {De Rijke}, Pavel Serdyukov, and Luo Si.
  2012.
\newblock \href {https://doi.org/10.1561/1500000024} {{Expertise retrieval}}.
\newblock \emph{Foundations and Trends in Information Retrieval},
  6(2-3):127--256.

\bibitem[{Beel et~al.(2016)Beel, Gipp, Langer, and Breitinger}]{Beel2016}
Joeran Beel, Bela Gipp, Stefan Langer, and Corinna Breitinger. 2016.
\newblock \href {https://doi.org/10.1007/s00799-015-0156-0} {{Research-paper
  recommender systems: a literature survey}}.
\newblock \emph{International Journal on Digital Libraries}, 17(4):305--338.

\bibitem[{Beltagy et~al.(2019)Beltagy, Lo, and Cohan}]{Beltagy2019SciBERT}
Iz~Beltagy, Kyle Lo, and Arman Cohan. 2019.
\newblock \href {https://www.aclweb.org/anthology/D19-1371.pdf} {Scibert:
  Pretrained language model for scientific text}.
\newblock In \emph{EMNLP}.

\bibitem[{Beltagy et~al.(2020)Beltagy, Peters, and Cohan}]{Beltagy2020}
Iz~Beltagy, Matthew~E. Peters, and Arman Cohan. 2020.
\newblock \href {http://arxiv.org/abs/2004.05150} {Longformer: The
  long-document transformer}.

\bibitem[{Berendsen et~al.(2013)Berendsen, De~Rijke, Balog, Bogers, and Van
  Den~Bosch}]{berendsen2013assessment}
Richard Berendsen, Maarten De~Rijke, Krisztian Balog, Toine Bogers, and Antal
  Van Den~Bosch. 2013.
\newblock On the assessment of expertise profiles.
\newblock \emph{Journal of the American Society for Information Science and
  Technology}, 64(10):2024--2044.

\bibitem[{Berger(2020)}]{markdataset2020}
Mark Berger. 2020.
\newblock \href {https://doi.org/10.5281/ZENODO.4075166} {Datasets for
  effective distributed representations for academic expert search}.

\bibitem[{Berger et~al.(2017)Berger, McDonough, and Seversky}]{Berger2017}
Matthew Berger, Katherine McDonough, and Lee~M. Seversky. 2017.
\newblock \href {https://doi.org/10.1109/TVCG.2016.2598667} {{Cite2vec:
  Citation-Driven Document Exploration via Word Embeddings}}.
\newblock \emph{IEEE Transactions on Visualization and Computer Graphics},
  23(1):691--700.

\bibitem[{Bin and Kuo(2020)}]{Bin2020}
Wang Bin and C.-C~Jay Kuo. 2020.
\newblock \href {http://arxiv.org/abs/2002.06652v1} {{SBERT-WK: A Sentence
  Embedding Method By Dissecting BERT-based Word Models}}.

\bibitem[{Bowman et~al.(2015)Bowman, Angeli, Potts, and Manning}]{Bowman2015}
Samuel~R. Bowman, Gabor Angeli, Christopher Potts, and Christopher~D. Manning.
  2015.
\newblock \href {https://doi.org/10.18653/v1/D15-1075} {{A large annotated
  corpus for learning natural language inference}}.
\newblock In \emph{Proceedings of the 2015 Conference on Empirical Methods in
  Natural Language Processing}, pages 632--642, Stroudsburg, PA, USA.
  Association for Computational Linguistics.

\bibitem[{Brochier et~al.(2018)Brochier, Guille, Rothan, and
  Velcin}]{Brochier2018}
Robin Brochier, Adrien Guille, Benjamin Rothan, and Julien Velcin. 2018.
\newblock \href {http://arxiv.org/abs/1806.10813} {{Impact of the query set on
  the evaluation of expert finding systems}}.
\newblock \emph{CEUR Workshop Proceedings}, 2132:32--45.

\bibitem[{Bromley et~al.(1994)Bromley, Guyon, LeCun, S{\"a}ckinger, and
  Shah}]{bromley1994signature}
Jane Bromley, Isabelle Guyon, Yann LeCun, Eduard S{\"a}ckinger, and Roopak
  Shah. 1994.
\newblock Signature verification using a" siamese" time delay neural network.
\newblock In \emph{Advances in neural information processing systems}, pages
  737--744.

\bibitem[{Cer et~al.()Cer, Diab, Agirre, {Nigo Lopez-Gazpio}, and Specia}]{Cer}
Daniel Cer, Mona Diab, Eneko Agirre, I~˜ {Nigo Lopez-Gazpio}, and Lucia
  Specia.
\newblock \href {http://www.sdl.com/languagecloud/} {{SemEval-2017 Task 1:
  Semantic Textual Similarity Multilingual and Cross-lingual Focused
  Evaluation}}.
\newblock pages 1--14.

\bibitem[{Cer et~al.(2018)Cer, Yang, Kong, Hua, Limtiaco, {St John}, Constant,
  Guajardo-C{\'{e}}spedes, Yuan, Tar, Sung, Strope, and {Kurzweil Google
  Research Mountain View}}]{Cer2018}
Daniel Cer, Yinfei Yang, Sheng-yi Kong, Nan Hua, Nicole Limtiaco, Rhomni {St
  John}, Noah Constant, Mario Guajardo-C{\'{e}}spedes, Steve Yuan, Chris Tar,
  Yun-Hsuan Sung, Brian Strope, and Ray {Kurzweil Google Research Mountain
  View}. 2018.
\newblock \href {http://arxiv.org/abs/1803.11175v2} {{Universal Sentence
  Encoder}}.

\bibitem[{Chen(2017)}]{chen2017efficient}
Minmin Chen. 2017.
\newblock \href {http://arxiv.org/abs/1707.02377} {Efficient vector
  representation for documents through corruption}.
\newblock \emph{CoRR}, abs/1707.02377.

\bibitem[{Cohan et~al.(2020)Cohan, Feldman, Beltagy, Downey, and
  Weld}]{specter_cohan_2020}
Arman Cohan, Sergey Feldman, Iz~Beltagy, Doug Downey, and Daniel Weld. 2020.
\newblock \href {https://doi.org/10.18653/v1/2020.acl-main.207} {Specter:
  Document-level representation learning using citation-informed transformers}.
\newblock pages 2270--2282.

\bibitem[{Conneau et~al.(2018)Conneau, Kiela, Schwenk, Barrault, and
  Bordes}]{Conneau2018}
Alexis Conneau, Douwe Kiela, Holger Schwenk, Loïc Barrault, and Antoine
  Bordes. 2018.
\newblock \href {http://arxiv.org/abs/1705.02364v5} {{Supervised Learning of
  Universal Sentence Representations from Natural Language Inference Data}}.

\bibitem[{Craswell et~al.(2005)Craswell, de~Vries, and
  Soboroff}]{craswell2005overview}
Nick Craswell, Arjen~P. de~Vries, and Ian Soboroff. 2005.
\newblock \href
  {https://trec.nist.gov/pubs/trec14/papers/old.overviews/ENTERPRISE.OVERVIEW.pdf}
  {Overview of the trec 2005 enterprise track}.
\newblock In \emph{TREC}.

\bibitem[{Deerwester et~al.(1990)Deerwester, Dumais, Furnas, Landauer, and
  Harshman}]{deerwester1990indexing}
Scott Deerwester, Susan~T Dumais, George~W Furnas, Thomas~K Landauer, and
  Richard Harshman. 1990.
\newblock \href
  {https://doi.org/https://doi.org/10.1002/(SICI)1097-4571(199009)41:6\%3C391::AID-ASI1\%3E3.0.CO;2-9}
  {Indexing by latent semantic analysis}.
\newblock \emph{Journal of the American society for information science},
  41(6):391--407.

\bibitem[{Dempster(2008)}]{dempster2008upper}
Arthur~P Dempster. 2008.
\newblock Upper and lower probabilities induced by a multivalued mapping.
\newblock In \emph{Classic works of the Dempster-Shafer theory of belief
  functions}, pages 57--72. Springer.

\bibitem[{Deng et~al.(2012)Deng, King, and Lyu}]{Deng2012}
Hongbo Deng, Irwin King, and Michael~R. Lyu. 2012.
\newblock \href {https://doi.org/10.1109/TSMCB.2011.2161980} {{Enhanced models
  for expertise retrieval using community-aware strategies}}.
\newblock \emph{IEEE Transactions on Systems, Man, and Cybernetics, Part B:
  Cybernetics}, 42(1):93--106.

\bibitem[{Devlin et~al.(2018)Devlin, Chang, Lee, and Toutanova}]{Devlin2018}
Jacob Devlin, Ming-Wei Chang, Kenton Lee, and Kristina Toutanova. 2018.
\newblock \href {http://arxiv.org/abs/1810.04805} {{BERT: Pre-training of Deep
  Bidirectional Transformers for Language Understanding}}.

\bibitem[{Diaz et~al.(2016)Diaz, Mitra, and Craswell}]{Diaz2016}
Fernando Diaz, Bhaskar Mitra, and Nick Craswell. 2016.
\newblock \href {https://doi.org/10.18653/v1/p16-1035} {{Query expansion with
  locally-trained word embeddings}}.
\newblock \emph{54th Annual Meeting of the Association for Computational
  Linguistics, ACL 2016 - Long Papers}, 1:367--377.

\bibitem[{F{\"{a}}rber(2019)}]{DBLP:conf/semweb/Farber19}
Michael F{\"{a}}rber. 2019.
\newblock \href {https://doi.org/10.1007/978-3-030-30796-7\_8} {{The Microsoft
  Academic Knowledge Graph: {A} Linked Data Source with 8 Billion Triples of
  Scholarly Data}}.
\newblock In \emph{{Proceedings of the 18th International Semantic Web
  Conference}}, {ISWC'19}, pages 113--129.

\bibitem[{Faruqui et~al.(2015)Faruqui, Dodge, Jauhar, Dyer, Hovy, and
  Smith}]{Faruqui2015}
Manaal Faruqui, Jesse Dodge, Sujay~K. Jauhar, Chris Dyer, Eduard Hovy, and
  Noah~A. Smith. 2015.
\newblock \href {https://doi.org/10.3115/v1/n15-1184} {{Retrofitting word
  vectors to semantic lexicons}}.
\newblock \emph{NAACL HLT 2015 - 2015 Conference of the North American Chapter
  of the Association for Computational Linguistics: Human Language
  Technologies, Proceedings of the Conference}, (i):1606--1615.

\bibitem[{Gon\c{c}alves and Dorneles(2019)}]{Goncalves2019}
Rodrigo Gon\c{c}alves and Carina~Friedrich Dorneles. 2019.
\newblock \href {https://doi.org/10.1145/3331000} {Automated expertise
  retrieval: A taxonomy-based survey and open issues}.
\newblock \emph{ACM Comput. Surv.}, 52(5).

\bibitem[{Gu and Blackmore(2016)}]{Gu2016}
Xin Gu and Karen~L. Blackmore. 2016.
\newblock \href {https://doi.org/10.1007/s11192-016-1985-3} {{Recent trends in
  academic journal growth}}.
\newblock \emph{Scientometrics}, 108(2):693--716.

\bibitem[{Harris(1954)}]{Harris1954}
Zellig~S. Harris. 1954.
\newblock \href {https://doi.org/10.1080/00437956.1954.11659520}
  {{Distributional Structure}}.
\newblock \emph{WORD}, 10(2-3):146--162.

\bibitem[{Hassan et~al.(2019{\natexlab{a}})Hassan, Sansonetti, Gasparetti,
  Micarelli, and Beel}]{Hassan2019}
Hebatallah A~Mohamed Hassan, Giuseppe Sansonetti, Fabio Gasparetti, Alessandro
  Micarelli, and Joeran Beel. 2019{\natexlab{a}}.
\newblock \href {https://tfhub.dev/google/universal-sentence-} {{BERT, ELMo,
  USE and InferSent Sentence Encoders: The Panacea for Research-Paper
  Recommendation?}}

\bibitem[{Hassan et~al.(2019{\natexlab{b}})Hassan, Sansonetti, Gasparetti,
  Micarelli, and Beel}]{hassan2019bert}
Hebatallah A~Mohamed Hassan, Giuseppe Sansonetti, Fabio Gasparetti, Alessandro
  Micarelli, and Joeran Beel. 2019{\natexlab{b}}.
\newblock Bert, elmo, use and infersent sentence encoders: The panacea for
  research-paper recommendation?
\newblock In \emph{RecSys (Late-Breaking Results)}, pages 6--10.

\bibitem[{Hill et~al.(2016)Hill, Cho, and
  Korhonen}]{hill-etal-2016-learning-distributed}
Felix Hill, Kyunghyun Cho, and Anna Korhonen. 2016.
\newblock \href {https://doi.org/10.18653/v1/N16-1162} {Learning distributed
  representations of sentences from unlabelled data}.
\newblock In \emph{Proceedings of the 2016 Conference of the North {A}merican
  Chapter of the Association for Computational Linguistics: Human Language
  Technologies}, pages 1367--1377, San Diego, California. Association for
  Computational Linguistics.

\bibitem[{Honnibal and Montani(2017)}]{spacy2}
Matthew Honnibal and Ines Montani. 2017.
\newblock {spaCy 2}: Natural language understanding with {B}loom embeddings,
  convolutional neural networks and incremental parsing.
\newblock To appear.

\bibitem[{Husain et~al.(2019)Husain, Salim, Alias, Abdelsalam, and
  Hassan}]{Husain2019}
Omayma Husain, Naomie Salim, Rose~Alinda Alias, Samah Abdelsalam, and Alzubair
  Hassan. 2019.
\newblock \href {https://doi.org/10.3390/app9204250} {{Expert finding systems:
  A systematic review}}.
\newblock \emph{Applied Sciences (Switzerland)}, 9(20):1--32.

\bibitem[{Ioannidis et~al.(2018)Ioannidis, Klavans, and Boyack}]{Ioannidis2018}
John~P.A. Ioannidis, Richard Klavans, and Kevin~W. Boyack. 2018.
\newblock \href {https://doi.org/10.1038/d41586-018-06185-8} {{Thousands of
  scientists publish a paper every five days}}.
\newblock \emph{Nature}, 561(7722):167--169.

\bibitem[{J et~al.(2016)J, Ganguly, Gupta, Varma, and Pudi}]{Soumyajit2015}
Ganesh J, Soumyajit Ganguly, Manish Gupta, Vasudeva Varma, and Vikram Pudi.
  2016.
\newblock \href {https://doi.org/10.1145/2872518.2889382} {Author2vec: Learning
  author representations by combining content and link information}.
\newblock In \emph{Proceedings of the 25th International Conference Companion
  on World Wide Web}, WWW ’16 Companion, page 49–50, Republic and Canton of
  Geneva, CHE. International World Wide Web Conferences Steering Committee.

\bibitem[{J{\"{a}}rvelin and Kek{\"{a}}l{\"{a}}inen(2002)}]{Jarvelin2002}
Kalervo J{\"{a}}rvelin and Jaana Kek{\"{a}}l{\"{a}}inen. 2002.
\newblock \href {https://doi.org/10.1145/582415.582418} {{Cumulated gain-based
  evaluation of IR techniques}}.
\newblock \emph{ACM Transactions on Information Systems (TOIS)},
  20(4):422--446.

\bibitem[{Johnson et~al.(2017)Johnson, Douze, and J{\'{e}}gou}]{JDH17}
Jeff Johnson, Matthijs Douze, and Herv{\'{e}} J{\'{e}}gou. 2017.
\newblock \href {http://arxiv.org/abs/1702.08734} {Billion-scale similarity
  search with gpus}.
\newblock \emph{CoRR}, abs/1702.08734.

\bibitem[{J{\'u}nior et~al.(2017)J{\'u}nior, Silva, Costa, and
  Amancio}]{Jnior2017PatternsOA}
E.~A.~C. J{\'u}nior, F.~N. Silva, L.~Costa, and Diego~R. Amancio. 2017.
\newblock \href {https://doi.org/10.1016/j.joi.2017.03.003} {Patterns of
  authors contribution in scientific manuscripts}.
\newblock \emph{ArXiv}, abs/1609.05545.

\bibitem[{Kenter et~al.(2016)Kenter, Borisov, and De~Rijke}]{kenter2016siamese}
Tom Kenter, Alexey Borisov, and Maarten De~Rijke. 2016.
\newblock Siamese cbow: Optimizing word embeddings for sentence
  representations.
\newblock \emph{arXiv preprint arXiv:1606.04640}.

\bibitem[{Le and Mikolov(2014)}]{Le2014}
Quoc~V. Le and Tomas Mikolov. 2014.
\newblock \href {http://arxiv.org/abs/1405.4053} {{Distributed Representations
  of Sentences and Documents}}.

\bibitem[{Lev et~al.(2015)Lev, Klein, and Wolf}]{lev2015defense}
Guy Lev, Benjamin Klein, and Lior Wolf. 2015.
\newblock In defense of word embedding for generic text representation.
\newblock In \emph{International Conference on Applications of Natural Language
  to Information Systems}, pages 35--50. Springer.

\bibitem[{Lin et~al.(2017)Lin, Hong, Wang, and Li}]{Lin2017}
Shuyi Lin, Wenxing Hong, Dingding Wang, and Tao Li. 2017.
\newblock \href {https://doi.org/10.1007/s10844-016-0440-5} {{A survey on
  expert finding techniques}}.
\newblock \emph{Journal of Intelligent Information Systems}, 49(2):255--279.

\bibitem[{Liu and Lapata(2017)}]{Liu_docemb}
Yang Liu and Mirella Lapata. 2017.
\newblock \href {http://arxiv.org/abs/1705.09207} {Learning structured text
  representations}.
\newblock \emph{CoRR}, abs/1705.09207.

\bibitem[{Liu et~al.(2019)Liu, Ott, Goyal, Du, Joshi, Chen, Levy, Lewis,
  Zettlemoyer, Stoyanov, and Allen}]{Liu2019}
Yinhan Liu, Myle Ott, Naman Goyal, Jingfei Du, Mandar Joshi, Danqi Chen, Omer
  Levy, Mike Lewis, Luke Zettlemoyer, Veselin Stoyanov, and Paul~G Allen. 2019.
\newblock \href {http://arxiv.org/abs/1907.11692v1} {{RoBERTa: A Robustly
  Optimized BERT Pretraining Approach}}.

\bibitem[{MacAvaney et~al.(2019)MacAvaney, Yates, Cohan, and
  Goharian}]{10.1145/3331184.3331317}
Sean MacAvaney, Andrew Yates, Arman Cohan, and Nazli Goharian. 2019.
\newblock \href {https://doi.org/10.1145/3331184.3331317} {Cedr: Contextualized
  embeddings for document ranking}.
\newblock In \emph{Proceedings of the 42nd International ACM SIGIR Conference
  on Research and Development in Information Retrieval}, SIGIR'19, page
  1101–1104, New York, NY, USA. Association for Computing Machinery.

\bibitem[{Macdonald(2009)}]{macdonald2009voting}
Craig Macdonald. 2009.
\newblock \emph{The voting model for people search}.
\newblock Ph.D. thesis, University of Glasgow.

\bibitem[{Macdonald and Ounis(2006{\natexlab{a}})}]{Macdonald2006_2}
Craig Macdonald and Iadh Ounis. 2006{\natexlab{a}}.
\newblock \href
  {http://www.dcs.gla.ac.uk/~craigm/publications/macdonald07networks.pdf} {{A
  Belief Network Model for Expert Search}}.
\newblock \emph{Computing}.

\bibitem[{Macdonald and Ounis(2006{\natexlab{b}})}]{Macdonald2006}
Craig Macdonald and Iadh Ounis. 2006{\natexlab{b}}.
\newblock \href {https://doi.org/10.1145/1183614.1183671} {{Voting for
  candidates}}.
\newblock In \emph{Proceedings of the 15th ACM international conference on
  Information and knowledge management - CIKM '06}, page 387, New York, New
  York, USA. ACM Press.

\bibitem[{Macdonald and Ounis(2008)}]{Macdonald2008}
Craig Macdonald and Iadh Ounis. 2008.
\newblock \href {https://doi.org/10.1007/s10115-007-0105-3} {{Voting techniques
  for expert search}}.
\newblock \emph{Knowl Inf Syst}, 16:259--280.

\bibitem[{Mai et~al.(2018)Mai, Janowicz, and Yan}]{mai2018combining}
Gengchen Mai, Krzysztof Janowicz, and Bo~Yan. 2018.
\newblock \href {http://ceur-ws.org/Vol-2241/paper-08.pdf} {Combining text
  embedding and knowledge graph embedding techniques for academic search
  engines}.
\newblock In \emph{Joint proceedings of the 4th Workshop on Semantic Deep
  Learning (SemDeep-4) and NLIWoD4: Natural Language Interfaces for the Web of
  Data {(NLIWOD-4)} and 9th Question Answering over Linked Data challenge
  {(QALD-9)} co-located with 17th International Semantic Web Conference {(ISWC}
  2018), Monterey, California, United States of America, October 8th - 9th,
  2018}, volume 2241 of \emph{{CEUR} Workshop Proceedings}, pages 77--88.
  CEUR-WS.org.

\bibitem[{{Malkov} and {Yashunin}(2020)}]{8594636}
Y.~A. {Malkov} and D.~A. {Yashunin}. 2020.
\newblock \href {https://doi.org/https://doi.org/10.1109/TPAMI.2018.2889473}
  {Efficient and robust approximate nearest neighbor search using hierarchical
  navigable small world graphs}.
\newblock \emph{IEEE Transactions on Pattern Analysis and Machine
  Intelligence}, 42(4):824--836.

\bibitem[{May et~al.(2019)May, Wang, Bordia, Bowman, and Rudinger}]{May2019}
Chandler May, Alex Wang, Shikha Bordia, Samuel~R. Bowman, and Rachel Rudinger.
  2019.
\newblock \href {http://arxiv.org/abs/1903.10561} {{On Measuring Social Biases
  in Sentence Encoders}}.

\bibitem[{Moreira and Wichert(2013)}]{Moreira2013}
Catarina Moreira and Andreas Wichert. 2013.
\newblock \href {https://doi.org/10.1016/j.eswa.2013.04.001} {{Finding academic
  experts on a multisensor approach using Shannon's entropy}}.
\newblock \emph{Expert Systems with Applications}, 40(14):5740--5754.

\bibitem[{Nikzad-Khasmakhi et~al.(2020)Nikzad-Khasmakhi, Balafar,
  Feizi-Derakhshi, and Motamed}]{Nikzad-Khasmakhi2020}
N.~Nikzad-Khasmakhi, M.~A. Balafar, M.~Reza Feizi-Derakhshi, and Cina Motamed.
  2020.
\newblock \href {http://arxiv.org/abs/2001.08503} {{ExEm: Expert Embedding
  using dominating set theory with deep learning approaches}}.
\newblock pages 1--27.

\bibitem[{Pagliardini et~al.(2018)Pagliardini, Gupta, and
  Jaggi}]{Pagliardini2017}
Matteo Pagliardini, Prakhar Gupta, and Martin Jaggi. 2018.
\newblock \href {https://doi.org/10.18653/v1/N18-1049} {Unsupervised learning
  of sentence embeddings using compositional n-gram features}.
\newblock In \emph{Proceedings of the 2018 Conference of the North {A}merican
  Chapter of the Association for Computational Linguistics: Human Language
  Technologies, Volume 1 (Long Papers)}, pages 528--540, New Orleans,
  Louisiana. Association for Computational Linguistics.

\bibitem[{Pedregosa et~al.(2011)Pedregosa, Varoquaux, Gramfort, Michel,
  Thirion, Grisel, Blondel, Prettenhofer, Weiss, Dubourg, Vanderplas, Passos,
  Cournapeau, Brucher, Perrot, and Duchesnay}]{scikit-learn}
F.~Pedregosa, G.~Varoquaux, A.~Gramfort, V.~Michel, B.~Thirion, O.~Grisel,
  M.~Blondel, P.~Prettenhofer, R.~Weiss, V.~Dubourg, J.~Vanderplas, A.~Passos,
  D.~Cournapeau, M.~Brucher, M.~Perrot, and E.~Duchesnay. 2011.
\newblock Scikit-learn: Machine learning in {P}ython.
\newblock \emph{Journal of Machine Learning Research}, 12:2825--2830.

\bibitem[{Pennington et~al.(2014)Pennington, Socher, and
  Manning}]{pennington2014glove}
Jeffrey Pennington, Richard Socher, and Christopher Manning. 2014.
\newblock \href {https://doi.org/10.3115/v1/D14-1162} {{G}lo{V}e: Global
  vectors for word representation}.
\newblock In \emph{Proceedings of the 2014 Conference on Empirical Methods in
  Natural Language Processing ({EMNLP})}, pages 1532--1543, Doha, Qatar.
  Association for Computational Linguistics.

\bibitem[{Price and Flach(2017)}]{price2017computational}
Simon Price and Peter~A Flach. 2017.
\newblock Computational support for academic peer review: A perspective from
  artificial intelligence.
\newblock \emph{Communications of the ACM}, 60(3):70--79.

\bibitem[{Reimers and Gurevych(2019)}]{Reimers2019}
Nils Reimers and Iryna Gurevych. 2019.
\newblock \href {https://doi.org/10.18653/v1/D19-1410} {Sentence-{BERT}:
  Sentence embeddings using {S}iamese {BERT}-networks}.
\newblock In \emph{Proceedings of the 2019 Conference on Empirical Methods in
  Natural Language Processing and the 9th International Joint Conference on
  Natural Language Processing (EMNLP-IJCNLP)}, pages 3982--3992, Hong Kong,
  China. Association for Computational Linguistics.

\bibitem[{Salton(1988)}]{10.5555/63601}
Gerald Salton. 1988.
\newblock \emph{Automatic Text Processing}.
\newblock Addison-Wesley Longman Publishing Co., Inc., USA.

\bibitem[{Sauermann and Haeussler(2017)}]{sauermann2017authorship}
Henry Sauermann and Carolin Haeussler. 2017.
\newblock Authorship and contribution disclosures.
\newblock \emph{Science Advances}, 3(11):e1700404.

\bibitem[{Schleyer et~al.(2012)Schleyer, Butler, Song, and
  Spallek}]{10.1145/2147783.2147785}
Titus Schleyer, Brian~S. Butler, Mei Song, and Heiko Spallek. 2012.
\newblock \href {https://doi.org/10.1145/2147783.2147785} {Conceptualizing and
  advancing research networking systems}.
\newblock \emph{ACM Trans. Comput.-Hum. Interact.}, 19(1).

\bibitem[{Serdyukov et~al.(2007)Serdyukov, Chernov, and Nejdl}]{Serdyukov2007}
Pavel Serdyukov, Sergey Chernov, and Wolfgang Nejdl. 2007.
\newblock {Query Modeling}.
\newblock pages 737--740.

\bibitem[{Shannon(1948)}]{Shannon1948}
C.~E. Shannon. 1948.
\newblock \href {https://doi.org/10.1002/j.1538-7305.1948.tb01338.x} {{A
  Mathematical Theory of Communication}}.
\newblock \emph{Bell System Technical Journal}, 27(3):379--423.

\bibitem[{Shibata et~al.(2008)Shibata, Kajikawa, Takeda, and
  Matsushima}]{SHIBATA2008758}
Naoki Shibata, Yuya Kajikawa, Yoshiyuki Takeda, and Katsumori Matsushima. 2008.
\newblock \href
  {https://doi.org/https://doi.org/10.1016/j.technovation.2008.03.009}
  {Detecting emerging research fronts based on topological measures in citation
  networks of scientific publications}.
\newblock \emph{Technovation}, 28(11):758 -- 775.

\bibitem[{Silva(2014)}]{silva2014research}
Attulugamage Thushari~Priyangika Silva. 2014.
\newblock \emph{A research analytics framework for expert recommendation in
  research social networks}.
\newblock Ph.D. thesis, City University of Hong Kong.

\bibitem[{Sinha et~al.()Sinha, Shen, Song, Ma, Eide, Hsu, and Wang}]{Sinha}
Arnab Sinha, Zhihong Shen, Yang Song, Hao Ma, Darrin Eide, Bo-june Hsu, and
  Kuansan Wang.
\newblock \href {https://doi.org/10.1145/2740908.2742839} {{An Overview of
  Microsoft Academic Service (MAS) and Applications}}.

\bibitem[{Sinoara et~al.(2019)Sinoara, Camacho-Collados, Rossi, Navigli, and
  Rezende}]{Sinoara2019}
Roberta~A Sinoara, Jose Camacho-Collados, Rafael~G Rossi, Roberto Navigli, and
  Solange~O Rezende. 2019.
\newblock \href {https://doi.org/10.1016/j.knosys.2018.10.026}
  {{Knowledge-enhanced document embeddings for text classification}}.
\newblock \emph{Knowledge-Based Systems}, 163:955--971.

\bibitem[{Sziklai(2018)}]{Sziklai2018HowTI}
Bal{\'a}zs Sziklai. 2018.
\newblock \href {https://doi.org/10.1007/s00182-017-0582-x} {How to identify
  experts in a community?}
\newblock \emph{International Journal of Game Theory}, 47:155--173.

\bibitem[{Tang et~al.(2008)Tang, Zhang, Yao, Li, Zhang, and Su}]{Tang2008}
Jie Tang, Jing Zhang, Limin Yao, Juanzi Li, Li~Zhang, and Zhong Su. 2008.
\newblock \href {https://doi.org/10.1145/1401890.1402008} {{ArnetMiner:
  Extraction and mining of academic social networks}}.
\newblock \emph{Proceedings of the ACM SIGKDD International Conference on
  Knowledge Discovery and Data Mining}, pages 990--998.

\bibitem[{Tho et~al.(2007)Tho, Hui, and Fong}]{THO2007248}
Quan~Thanh Tho, S.C. Hui, and A.C.M. Fong. 2007.
\newblock \href {https://doi.org/https://doi.org/10.1016/j.ipm.2006.05.015} {A
  citation-based document retrieval system for finding research expertise}.
\newblock \emph{Information Processing \& Management}, 43(1):248 -- 264.

\bibitem[{Vrande\v{c}i\'{c} and Kr\"{o}tzsch(2014)}]{vrandevcic2014wikidata}
Denny Vrande\v{c}i\'{c} and Markus Kr\"{o}tzsch. 2014.
\newblock \href {https://doi.org/10.1145/2629489} {Wikidata: A free
  collaborative knowledgebase}.
\newblock \emph{Commun. ACM}, 57(10):78–85.

\bibitem[{Wang et~al.(2020{\natexlab{a}})Wang, Shen, Huang, Wu, Dong, and
  Kanakia}]{wang2020microsoft}
Kuansan Wang, Zhihong Shen, Chiyuan Huang, Chieh-Han Wu, Yuxiao Dong, and
  Anshul Kanakia. 2020{\natexlab{a}}.
\newblock \href {https://doi.org/https://doi.org/10.1162/qss_a_00021}
  {Microsoft academic graph: When experts are not enough}.
\newblock \emph{Quantitative Science Studies}, 1(1):396--413.

\bibitem[{Wang et~al.(2016)Wang, Tang, Aggarwal, and Liu}]{wang2016linked}
Suhang Wang, Jiliang Tang, Charu Aggarwal, and Huan Liu. 2016.
\newblock \href {https://doi.org/10.1145/2983323.2983755} {Linked document
  embedding for classification}.
\newblock In \emph{Proceedings of the 25th ACM International on Conference on
  Information and Knowledge Management}, CIKM ’16, page 115–124, New York,
  NY, USA. Association for Computing Machinery.

\bibitem[{Wang et~al.(2020{\natexlab{b}})Wang, Hou, Che, and Liu}]{Wang2020}
Yuxuan Wang, Yutai Hou, Wanxiang Che, and Ting Liu. 2020{\natexlab{b}}.
\newblock \href {https://doi.org/10.1007/s13042-020-01069-8} {{From static to
  dynamic word representations: a survey}}.
\newblock \emph{International Journal of Machine Learning and Cybernetics}.

\bibitem[{Wesley-Tanaskovic et~al.(1994)Wesley-Tanaskovic, Tocatlian, and
  Roberts}]{wesley1994expanding}
Ines Wesley-Tanaskovic, Jacques Tocatlian, and Kenneth~H Roberts. 1994.
\newblock \emph{Expanding Access to Science and Technology: The Role of
  Information Technologies: Proceedings of the Second International Symposium
  on the Frontiers of Science and Technology Held in Kyoto, Japan, 12-14 May
  1992}.
\newblock United Nations University Press.

\bibitem[{Williams et~al.(2018)Williams, Nangia, and Bowman}]{Williams2018}
Adina Williams, Nikita Nangia, and Samuel Bowman. 2018.
\newblock \href {https://doi.org/10.18653/v1/N18-1101} {{A Broad-Coverage
  Challenge Corpus for Sentence Understanding through Inference}}.
\newblock In \emph{Proceedings of the 2018 Conference of the North American
  Chapter of the Association for Computational Linguistics: Human Language
  Technologies, Volume 1 (Long Papers)}, pages 1112--1122, Stroudsburg, PA,
  USA. Association for Computational Linguistics.

\bibitem[{Yan et~al.(2011)Yan, Tang, Liu, Shan, and
  Li}]{10.1145/2063576.2063757}
Rui Yan, Jie Tang, Xiaobing Liu, Dongdong Shan, and Xiaoming Li. 2011.
\newblock \href {https://doi.org/10.1145/2063576.2063757} {Citation count
  prediction: Learning to estimate future citations for literature}.
\newblock In \emph{Proceedings of the 20th ACM International Conference on
  Information and Knowledge Management}, CIKM '11, page 1247–1252, New York,
  NY, USA. Association for Computing Machinery.

\bibitem[{Zhan et~al.(2011)Zhan, Yang, Bao, Han, Su, and
  Yu}]{10.5555/2035562.2035601}
Zhenjiang Zhan, Lichun Yang, Shenghua Bao, Dingyi Han, Zhong Su, and Yong Yu.
  2011.
\newblock \href {https://doi.org/10.1007/978-3-642-23535-1_29} {Finding
  appropriate experts for collaboration}.
\newblock In \emph{Proceedings of the 12th International Conference on Web-Age
  Information Management}, WAIM'11, page 327–339, Berlin, Heidelberg.
  Springer-Verlag.

\bibitem[{Zhang et~al.(2019{\natexlab{a}})Zhang, Kishore, Wu, Weinberger, and
  Artzi}]{Zhang2019bert}
Tianyi Zhang, Varsha Kishore, Felix Wu, Kilian~Q. Weinberger, and Yoav Artzi.
  2019{\natexlab{a}}.
\newblock \href {http://arxiv.org/abs/1904.09675} {{BERTScore: Evaluating Text
  Generation with BERT}}.

\bibitem[{Zhang et~al.(2016)Zhang, Rahman, Braylan, Dang, Chang, Kim, McNamara,
  Angert, Banner, Khetan, McDonnell, Nguyen, Xu, Wallace, and
  Lease}]{Zhang2016}
Ye~Zhang, Md~Mustafizur Rahman, Alex Braylan, Brandon Dang, Heng-Lu Chang,
  Henna Kim, Quinten McNamara, Aaron Angert, Edward Banner, Vivek Khetan, Tyler
  McDonnell, An~Thanh Nguyen, Dan Xu, Byron~C. Wallace, and Matthew Lease.
  2016.
\newblock \href {http://arxiv.org/abs/1611.06792} {{Neural Information
  Retrieval: A Literature Review}}.
\newblock (November).

\bibitem[{Zhang(2019)}]{Zhang2019retro}
Yi~Zhang. 2019.
\newblock \emph{{Learning Embeddings for Academic Papers}}.
\newblock Ph.D. thesis.

\bibitem[{Zhang et~al.(2019{\natexlab{b}})Zhang, Zhao, and Lu}]{Zhang2019}
Yi~Zhang, Fen Zhao, and Jianguo Lu. 2019{\natexlab{b}}.
\newblock \href {https://doi.org/10.1007/s11192-019-03206-9} {{P2V: large-scale
  academic paper embedding}}.
\newblock \emph{Scientometrics}, 121(1):399--432.

\bibitem[{Zhu et~al.(2009)Zhu, Song, and R{\"{u}}ger}]{Zhu2009}
Jianhan Zhu, Dawei Song, and Stefan R{\"{u}}ger. 2009.
\newblock \href {https://doi.org/10.1002/asi.21012} {{Integrating multiple
  windows and document features for expert finding}}.
\newblock \emph{Journal of the American Society for Information Science and
  Technology}, 60(4):694--715.

\end{thebibliography}
\bibliographystyle{acl_natbib}

% \clearpage
% \section{Algorithms}
% \appendix

% \section{Algorithms}

% \label{sec:appendix}
\clearpage
\appendix
% \section{Algorithms}

\begin{algorithm*}[hpb]
\DontPrintSemicolon
\SetAlgoLined
\SetNoFillComment
\KwResult{Paper embeddings $ME$ following the \textbf{merge} strategy}
\smallbreak
 \KwIn{A set of paper titles $T$, a set of batches of abstract sentences $A$ (with $|T| = |A|$) and an embedder model $Embedder$}
 \bigbreak
 mergedEmbeddings $\gets []$\;
 abstractEmbeddingBatches $\gets []$\;
 
 titleEmbeddings $\gets Embedder.embed(T)$ \;
 
 \For{aBatch $\in$ A}
 {
 batchEmbeddings $\gets Embedder.embed(aBatch)$\;
 abstractEmbeddingBatches.\textit{append}(batchEmbeddings)\;
 }
 \textbf{assert} $|titleEmbeddings| = |abstractEmbeddingBatches|$\;
 \For{tE $\in$ titleEmbeddings, \hspace{0.1cm} aEB $\in$ abstractEmbeddingBatches}
 {
 aEB\textit{.append}(tE)\;
 N $\gets dim(aEB)$ \;
 mergedEmbedding $\gets \frac{1}{N}\sum\limits_{i=1}^{N}(emb_i \in aEB)$\Comment*[r]{\smaller Take the element-wise average of all the embeddings, resulting in one embedding}
 mergedEmbeddings.\textit{append}(mergedEmbedding)
 }
 
 \Return mergedEmbeddings

 \caption{Paper embedding creation following the merge strategy.}
 \label{merged-strategy}
\end{algorithm*}

\begin{algorithm*}[hpb]
\DontPrintSemicolon
\SetAlgoLined
\SetNoFillComment
\KwResult{Baseline paper embeddings $ME$ created by either BERT or GloVe embedding model.}
\smallbreak
 \KwIn{A set of batches of abstract sentences $A$ with elementwise appended corresponding paper titles $T$ (with $|T| = |A|$) and an embedder model $Embedder$ (either BERT or GloVe)}
 \bigbreak
 embeddings $\gets []$\;
 abstractEmbeddingBatches $\gets []$\;
 
 \For{aBatch $\in$ A}
 {
 batchEmbeddings $\gets Embedder.embed(aBatch)$\;
 abstractEmbeddingBatches.\textit{append}(batchEmbeddings)\;
 }

 \For{aEB $\in$ abstractEmbeddingBatches}
 {
 N $\gets dim(aEB)$ \;
 pooledEmbedding $\gets \frac{1}{N}\sum\limits_{i=1}^{N}(emb_i \in aEB)$\Comment*[r]{\small{Take the element-wise average of all the embeddings, resulting in one embedding}}
 embeddings.\textit{append}(pooledEmbedding)
 }
 
 \Return embeddings

 \caption{Paper embedding creation following baseline BERT or GloVe strategy.}
 \label{baseline-bert-glove-strategy}
\end{algorithm*}

\begin{algorithm*}[hbp]
\DontPrintSemicolon
\SetAlgoLined
\SetNoFillComment
\KwResult{Paper embeddings $ME$ following the \textbf{separate} strategy}
\smallbreak
 \KwIn{A set of paper titles $T$, a set of batches of abstract sentences $A$ (with $|T| = |A|$) and an embedder model $Embedder$}
 \bigbreak
 mergedEmbeddings $\gets []$\;
 abstractAverageEmbeddings $\gets []$\;
 
 titleEmbeddings $\gets Embedder.embed(T)$ \;
 
 \For{aBatch $\in$ A}
 {
 batchEmbeddings $\gets Embedder.embed(aBatch)$\;
 N $\gets dim(batchEmbeddings)$\;
 averageBatchEmbedding $\gets \frac{1}{N}\sum\limits_{i=1}^{N}(emb_i \in batchEmbeddings)$\Comment*[r]{\smaller Create an average embedding for abstract sentences only.}
 abstractAverageEmbeddings.\textit{append}(averageBatchEmbedding)\;
 }
 \textbf{assert} $|titleEmbeddings| = |abstractAverageEmbeddings|$\;
 \For{tE $\in$ titleEmbeddings, \hspace{0.1cm} aE $\in$ abstractAverageEmbeddings}
 {
 separateArray $\gets [tE, aE]$\Comment*[r]{\smaller Create a two item array consisting of the title embedding and the average abstract embedding.}
 N $\gets dim(separateArray)$\;
 separateEmbedding $\gets \frac{1}{N}\sum\limits_{i=1}^{N}(emb_i \in separateArray)$\Comment*[r]{\small Create an average embedding over just two embeddings.}
 separateEmbeddings.\textit{append}(separateEmbedding)
 }
 
 \Return separateEmbeddings

 \caption{Paper embedding creation following the separate strategy.}
 \label{separate-strategy}
\end{algorithm*}

\begin{algorithm*}[hbp]
\DontPrintSemicolon
\SetAlgoLined
\SetNoFillComment
\SetKwProg{Retrofit}{Retrofit}{}{}

\KwResult{Retrofitted paper embeddings }

\smallbreak

 \KwData{A mapping between paper id's and their embeddings $\rightarrow$ \textbf{\textit{C}} (corpus), the mapping between paper id's in lexicon and their references $\rightarrow$ \textbf{\textit{L}} (lexicon), and the amount of algorithm iterations $\rightarrow$ \textbf{\textit{numIter}}.}
 
 \bigbreak
 
 \Retrofit{$(C, L, numIter)$}{

 \textit{newCorpus} $\gets$ deepcopy (\textit{C}) \Comment*[r]{\small{Deep copy the corpus into a new variable to alter}}
 
 \textit{I} $\gets$ \hspace{0.02cm}\textit{newCorpus}.keys()\hspace{0.02cm} \Comment*[r]{\small{Extract all the keys (paper id's) from the corpus}}
 
 \textit{corpusVocabulary} $\gets$ set (\textit{I}) \Comment*[r]{\small{Consider only the unique paper id's}}
 
 \smallbreak
 
 \textit{\textcolor{blue}{// Two lines below exist for the case where the paper id's in lexicon differ from paper id's in corpus, normally not the case in our environment. Otherwise, they have not effect.}}\;
 \textit{LI} $\gets$ \hspace{0.02cm}\textit{L}.keys()\hspace{0.02cm} \Comment*[r]{\tiny{Extract all the keys (paper id's) from the lexicon}}
 \textit{relevantVocabulary} $\gets$ \textit{corpusVocabulary} $\cap$ set (\textit{LI}) \Comment*[r]{\small{Consider only the overlapping id's from lexicon and corpus}}
 
 \For{$it \gets 0$ to $numIter$} 
 {
    \ForEach{paper $\in$ relevantVocabulary}
    {
    $paperNeighbours \gets$ set $(L[word]) \cap corpusVocabulary$ \Comment*[r]{\tiny{Extract the set of neighbours of the current paper that actually have an embedding in our corpus}}
    $numNeighbours \gets$ len ($paperNeighbours)$\;
    \If{numNeighbours = 0}{ 
    \textbf{continue}\Comment*[r]{\tiny{If the paper has no neighbours, do not adjust the embedding and go to next paper}}}
    $newEmbedding \gets numNeighbours * C[paper]$ \Comment*[r]{\tiny{Initialize the new embedding by weighing in the original embedding}}
    \ForEach{neighbour $in$ paperNeighbours}{
    $newEmbedding  \mathrel{+}= newCorpus[neighbour]$ \Comment*[r]{\tiny{Add the neighbour embedding with weight 1.}}
    }
    $newCorpus[paper] \gets \frac{newEmbedding}{2 * numNeighbours}$ \Comment*[r]{\tiny{Finalize the new embedding by dividing the current new embedding by (2 * number\_neighbours), essentially putting the embedding back in the same space.}}
        }
    }
    \Return \textit{newCorpus}
 }
%  \While{While condition}{
%   instructions\;
%   \eIf{condition}{ 
%   instructions1 \Comment*[f]{Some comment}\;
%   instructions2\;
%   }{
%   instructions3\;
%   }
%  }

\caption{The retrofitting algorithm.}
\label{retro-algo-figure}
\end{algorithm*}

% \begin{algorithm*}[hbp]
% \DontPrintSemicolon
% \SetAlgoLined
% \SetNoFillComment
% \KwResult{Lexicon and corpus building for paper embedding retrofitting }
% \smallbreak
%  \KwData{A set of paper id's \textbf{$P$}, their corresponding embeddings \textbf{$\hat{Q}$} and the list of the corresponding paper id's they reference \textbf{$p_{cits}$}}
%  \bigbreak
%  lexicon $\gets \{ \}$ \Comment*[r]{\small{Initialize corpus and lexicon to empty dictionaries}}
%  corpus $\gets \{ \}$\;
%  \For{\textbf{$p_i$} $\in$ $P$, \hspace{0.1cm} \textbf{$pc_i$} $\in$ $p_{cits}$, \hspace{0.1cm} \textbf{$\hat{q_i}$} $\in$ $\hat{Q}$\Comment*[r]{\small{Loop in parallel (Python \textit{zip} structure)}}}
%  {
%     lexicon[\textbf{$p_i$}] $\gets$ \textbf{$pc_i$} \Comment*[r]{\tiny{Assign the references to a certain paper (can be an empty list)}}\;
%     corpus[\textbf{$p_i$}] $\gets$ \(\frac{\hat{q_i}}
%     {\sqrt{\sum\limits_{j=1}^{n}(\hat{q_i}^2)+ 1e-6}}\) \Comment*[r]{\small{Normalize and store the embedding}}\;
    
%     % embedding / math.sqrt((embedding**2).sum() + 1e-6)
%  }
%  \Return lexicon, corpus
% %  \While{While condition}{
% %   instructions\;
% %   \eIf{condition}{ 
% %   instructions1 \Comment*[f]{Some comment}\;
% %   instructions2\;
% %   }{
% %   instructions3\;
% %   }
% %  }
%  \caption{Lexicon and corpus creation for the retrofitting algorithm.}
%  \label{retro-lexicon-creation}
% \end{algorithm*}

\clearpage

\begin{algorithm*}[hpb]
\DontPrintSemicolon
\SetAlgoLined
\SetNoFillComment
\KwResult{A stratified sample of authors $randomAuthorSample$ of size $s=5000$. }
\smallbreak
 \KwIn{A set of authors $authorsArxiv$ extracted from the the arXiv subset with the corresponding metadata and the final sample size $s=5000$ (arbitrarily chosen).}
 \bigbreak
 \begin{enumerate}
    \item Filter $authorsArxiv$ to only contain authors which, for their papers, have references in the data. That is needed to be able to do retrofitting later.
    \item Given this new proper set of authors, $properA$, initialize four bins (strata) based on the amount of author publications:
        \begin{enumerate}
            \item 5 - 10 publications
            \item 10 - 50 publications
            \item 50 - 100 publications
            \item 100+ publications
        \end{enumerate}
    \item Calculate the bin size for each bin and also the total amount of authors in all the bins. 
    \item Perform proportionate allocation using a sampling fraction in each of the strata that is proportional to that of the total population. We went for a sample size of 5000 authors. For a single bin, the allocation process is as following:
    \smallbreak
    \textbf{Bin 1 (5 - 10 publications)} has 22.943 authors. The total population of authors in all the bins is 35.450 authors ($|properA|$). The share of Bin1 in the final 5.000 authors set is then

    \textbf{Bin 1} $=>$ 22.943 * (5.000 / 35.450) = 3.235 authors
    \item Proceed doing a simple random sampling from those pools of authors, resulting in \textit{randomAuthorSample}.
 \end{enumerate}

 \Return \textit{randomAuthorSample}

 \caption{The author sampling strategy from the arXiv subset.}
 \label{sampling-strategy}
\end{algorithm*}

% \usepackage{graphicx}
\begin{table*}[!h]
\centering
\resizebox{\textwidth}{!}{%
\begin{tabular}{llllllllllll}
\hline
\textbf{PAPER EMBEDDING KIND} &
  \textbf{DATA FUSION TECHNIQUE} &
  \textbf{\begin{tabular}[c]{@{}l@{}}MRR@10 \\ EXACT\end{tabular}} &
  \textbf{\begin{tabular}[c]{@{}l@{}}MRR@10 \\ APPRX\end{tabular}} &
  \textbf{\begin{tabular}[c]{@{}l@{}}MAP@10 \\ EXACT\end{tabular}} &
  \textbf{\begin{tabular}[c]{@{}l@{}}MAP@10 \\ APPRX\end{tabular}} &
  \textbf{\begin{tabular}[c]{@{}l@{}}MP@10\\ EXACT\end{tabular}} &
  \textbf{\begin{tabular}[c]{@{}l@{}}MP@10\\ APPRX\end{tabular}} &
  \textbf{\begin{tabular}[c]{@{}l@{}}MP@5 \\ EXACT\end{tabular}} &
  \textbf{\begin{tabular}[c]{@{}l@{}}MP@5 \\ APPRX\end{tabular}} &
  \textbf{NDCG@10} &
  \textbf{NDCG@5} \\ \hline
 LSI &
  $expCombSUM_{uniform}$ &
  0.75 &
  0.864 &
  0.399 &
  0.603 &
  0.462 &
  0.662 &
  0.496 &
  0.696 &
  0.39 &
  0.42 \\ \hline
 &
  $expCombSUM_{binary}$ &
  0.753 &
  0.844 &
  0.428 &
  0.626 &
  0.491 &
  0.693 &
  0.512 &
  0.708 &
  0.41 &
  0.44 \\ \hline
 &
  $expCombSUM_{descending}$ &
  0.755 &
  0.857 &
  0.411 &
  0.603 &
  0.476 &
  0.668 &
  0.484 &
  0.684 &
  0.39 &
  0.42 \\ \hline
 &
  $expCombSUM_{parabolic}$ &
  0.763 &
  0.864 &
  0.392 &
  0.587 &
  0.457 &
  0.651 &
  0.482 &
  0.68 &
  0.39 &
  0.42 \\ \hline
Average pooled BERT &
  $expCombSUM_{uniform}$ &
  0.554 &
  - &
  0.117 &
  - &
  0.198 &
  - &
  0.206 &
  - &
  0.1 &
  0.11 \\ \hline
 &
  $expCombSUM_{binary}$ &
  0.558 &
  - &
  0.12 &
  - &
  0.203 &
  - &
  0.222 &
  - &
  0.12 &
  0.13 \\ \hline
 &
  $expCombSUM_{descending}$ &
  0.556 &
  - &
  0.117 &
  - &
  0.2 &
  - &
  0.208 &
  - &
  0.11 &
  0.12 \\ \hline
 &
  $expCombSUM_{parabolic}$ &
  0.56 &
  - &
  0.107 &
  - &
  0.18 &
  - &
  0.218 &
  - &
  0.1 &
  0.12 \\ \hline
\begin{tabular}[c]{@{}l@{}}Average pooled GloVe\end{tabular} &
  $expCombSUM_{uniform}$ &
  0.626 &
  - &
  0.272 &
  - &
  0.369 &
  - &
  0.398 &
  - &
  0.27 &
  0.29 \\ \hline
 &
  $expCombSUM_{binary}$ &
  0.672 &
  - &
  0.304 &
  - &
  0.402 &
  - &
  0.414 &
  - &
  0.3 &
  0.31 \\ \hline
 &
  $expCombSUM_{descending}$ &
  0.661 &
  - &
  0.286 &
  - &
  0.383 &
  - &
  0.398 &
  - &
  0.28 &
  0.3 \\ \hline
 &
  $expCombSUM_{parabolic}$ &
  0.676 &
  - &
  0.254 &
  - &
  0.33 &
  - &
  0.392 &
  - &
  0.25 &
  0.28 \\ \hline
Merged Sentence-BERT &
  $expCombSUM_{uniform}$ &
  0.834 &
  \textbf{0.903} &
  0.419 &
  0.586 &
  0.491 &
  0.658 &
  0.528 &
  0.684 &
  0.43 &
  0.47 \\ \hline
 &
  $expCombSUM_{binary}$ &
  0.83 &
  0.893 &
  0.437 &
  0.606 &
  0.509 &
  0.677 &
  0.546 &
  0.708 &
  0.42 &
  0.46 \\ \hline
 &
  $expCombSUM_{descending}$ &
  0.818 &
  \textbf{0.903} &
  0.419 &
  0.59 &
  0.493 &
  0.657 &
  0.526 &
  0.688 &
  0.41 &
  0.46 \\ \hline
 &
  $expCombSUM_{parabolic}$ &
  0.812 &
  0.869 &
  0.4 &
  0.555 &
  0.484 &
  0.634 &
  0.508 &
  0.666 &
  0.41 &
  0.45 \\ \hline
\textbf{Separate Sentence-BERT} &
  $expCombSUM_{uniform}$ &
  0.838 &
  0.875 &
  0.495 &
  0.673 &
  0.572 &
  0.744 &
  0.616 &
  0.764 &
  0.53 &
  0.58 \\ \hline
 &
  $\mathbf{expCombSUM_{binary}}$ &
  0.837 &
  0.892 &
  0.518 &
  \textbf{0.69} &
  \textbf{0.59} &
  \textbf{0.751} &
  \textbf{0.626} &
  \textbf{0.784} &
  \textbf{0.54} &
  \textbf{0.6} \\ \hline
 &
  \begin{tabular}[c]{@{}l@{}}$Norm(expCombSUM_{binary})$\\ $\beta =0$ and $\alpha=1$\end{tabular} &
  0.619 &
  0.678 &
  0.174 &
  0.284 &
  0.282 &
  0.42 &
  0.272 &
  0.398 &
  0.15 &
  0.14 \\ \hline
 &
  \begin{tabular}[c]{@{}l@{}}$Norm(expCombSUM_{binary})$\\ $\beta =0$ and $\alpha=1000$\end{tabular} &
  0.694 &
  0.737 &
  0.218 &
  0.331 &
  0.318 &
  0.452 &
  0.324 &
  0.456 &
  0.16 &
  0.17 \\ \hline
 &
  \begin{tabular}[c]{@{}l@{}}$Norm(expCombSUM_{binary})$\\ $\beta =10$ and $\alpha=1000$\end{tabular} &
  0.777 &
  0.807 &
  0.293 &
  0.409 &
  0.381 &
  0.509 &
  0.406 &
  0.542 &
  0.22 &
  0.25 \\ \hline
 &
  \begin{tabular}[c]{@{}l@{}}$Norm(expCombSUM_{binary})$\\ $\beta = 50$ and $\alpha=1000$\end{tabular} &
  0.769 &
  0.81 &
  0.362 &
  0.49 &
  0.455 &
  0.589 &
  0.466 &
  0.602 &
  0.31 &
  0.33 \\ \hline
 &
  \begin{tabular}[c]{@{}l@{}}$Norm(expCombSUM_{binary})$\\ $\beta =1000$ and $\alpha=1000$\end{tabular} &
  0.813 &
  0.853 &
  0.404 &
  0.548 &
  0.491 &
  0.638 &
  0.52 &
  0.658 &
  0.37 &
  0.4 \\ \hline
 &
  $expCombSUM_{descending}$ &
  0.839 &
  0.894 &
  0.501 &
  0.666 &
  0.581 &
  0.737 &
  0.612 &
  0.758 &
  0.52 &
  0.58 \\ \hline
 &
  $expCombSUM_{parabolic}$ &
  0.819 &
  0.878 &
  0.486 &
  0.654 &
  0.565 &
  0.729 &
  0.592 &
  0.742 &
  0.52 &
  0.56 \\ \hline
Retrofitted merged Sentence-BERT &
  $expCombSUM_{uniform}$ &
  0.792 &
  0.839 &
  0.384 &
  0.551 &
  0.45 &
  0.61 &
  0.482 &
  0.638 &
  0.38 &
  0.42 \\ \hline
 &
  $expCombSUM_{binary}$ &
  0.83 &
  0.865 &
  0.404 &
  0.563 &
  0.467 &
  0.618 &
  0.496 &
  0.652 &
  0.39 &
  0.44 \\ \hline
 &
  $expCombSUM_{descending}$ &
  0.813 &
  0.843 &
  0.39 &
  0.551 &
  0.454 &
  0.609 &
  0.486 &
  0.644 &
  0.38 &
  0.42 \\ \hline
 &
  $expCombSUM_{parabolic}$ &
  0.775 &
  0.847 &
  0.38 &
  0.542 &
  0.445 &
  0.598 &
  0.474 &
  0.638 &
  0.38 &
  0.4 \\ \hline
Retrofitted separate Sentence-BERT &
  $expCombSUM_{uniform}$ &
  0.821 &
  0.893 &
  0.51 &
  0.658 &
  0.577 &
  0.716 &
  0.606 &
  0.732 &
  0.5 &
  0.54 \\ \hline
 &
  $expCombSUM_{binary}$ &
  \textbf{0.841} &
  0.893 &
  \textbf{0.519} &
  0.661 &
  0.584 &
  0.718 &
  0.616 &
  0.75 &
  0.51 &
  0.54 \\ \hline
 &
  $expCombSUM_{descending}$ &
  0.831 &
  0.895 &
  0.505 &
  0.647 &
  0.569 &
  0.702 &
  0.61 &
  0.734 &
  0.49 &
  0.54 \\ \hline
 &
  $expCombSUM_{parabolic}$ &
  0.808 &
  0.863 &
  0.509 &
  0.658 &
  0.583 &
  0.724 &
  0.596 &
  0.732 &
  0.5 &
  0.53 \\ \hline
\end{tabular}%
}
\caption{\small{Results for the voting model author retrieval strategy. The best results are formatted in bold.}}
\label{tab:results_voting_author_retrieval_full}
\end{table*}

% \usepackage{booktabs}
% \usepackage{graphicx}
\begin{table*}[!hbtp]
\centering
\resizebox{\textwidth}{!}{%
\begin{tabular}{@{}|l|l|l|l|@{}}
\toprule
\multicolumn{4}{|c|}{Queries}                                                                               \\ \midrule
'cluster analysis'              & 'Bayesian statistics'         & 'world wide web'                  & 'Novelty detection'           \\ \midrule
'Image segmentation'            & 'kernel density estimation'   & 'gibbs sampling'                  & 'semantic grid'               \\ \midrule
'Parallel algorithm'            & 'learning to rank'            & 'user interface'                  & 'Knowledge extraction'        \\ \midrule
'Monte Carlo method'            & 'relational database'         & 'belief propagation'              & 'Computational biology'       \\ \midrule
'Convex optimization'           & 'activity recognition'        & 'interpolation'                   & 'Web 2.0'                     \\ \midrule
'Dimensionality reduction'      & 'wearable computer'           & 'wavelet transform'               & 'Network theory'              \\ \midrule
'Facial recognition system'     & 'ensemble learning'           & 'transfer of learning'            & 'Video denoising'             \\ \midrule
'k-nearest neighbors algorithm' & 'wordnet'                     & 'topic model'                     & 'Quantum information science' \\ \midrule
'Hierarchical clustering'   & 'medical imaging'                  & 'clustering high-dimensional data' & 'Color quantization'         \\ \midrule
'Automatic text summarization'  & 'deconvolution'               & 'game theory'                     & 'social web'                  \\ \midrule
'Dynamic programming'           & 'Latent Dirichlet allocation' & 'biometrics'                      & 'entity linking'              \\ \midrule
'Genetic algorithm'             & 'Euclidian distance'          & 'constraint satisfaction'         & 'information privacy'         \\ \midrule
'Human-computer interaction'    & 'web service'                 & 'combinatorial optimization'      & 'random forest'               \\ \midrule
'Categorial grammar'            & 'multi-task learning'         & 'speech processing'               & 'cloud computing'             \\ \midrule
'Semantic Web'                  & 'Linear separability'         & 'multi-agent system'              & 'Knapsack problem'            \\ \midrule
'fuzzy logic'                   & 'OWL-S'                       & 'mean field theory'               & 'Linear algebra'              \\ \midrule
'image restoration'             & 'Wireless sensor network'     & 'social network'                  & 'batch processing'            \\ \midrule
'generative model'              & 'Semantic role labeling'      & 'lattice model'                   & 'rule induction'              \\ \midrule
'search algorithm'          & 'Continuous-time Markov chain'     & 'automatic image annotation'       & 'Uncertainty quantification' \\ \midrule
'sample size determination' & 'Open Knowledge Base Connectivity' & 'computational geometry'           & 'Computer architecture'      \\ \midrule
'anomaly detection'         & 'Propagation of uncertainty'       & 'Evolutionary algorithm'           & 'Best-first search'          \\ \midrule
'sentiment analysis'            & 'Fast Fourier transform'      & 'web search query'                & 'Gaussian random field'       \\ \midrule
'semantic similarity'           & 'Security token'              & 'eye tracking'                    & 'Support vector machine'      \\ \midrule
'logic programming'             & 'machine translation'         & 'query optimization'              & 'ontology language'           \\ \midrule
'Hyperspectral imaging'         & 'middleware'                  & 'Newton's method' & 'big data'                    \\ \bottomrule
\end{tabular}%
}
\caption{The full test queries set used for the system evaluation.}
\label{tab:test_queries}
\end{table*}

\end{document}